\RequirePackage[2020-02-02]{latexrelease}
\documentclass[aps,prb,twocolumn,floatfix,groupedaddress]
 {revtex4}
\usepackage[T1]{fontenc}
\usepackage[latin9]{inputenc}
\usepackage{graphicx}
\usepackage{amssymb}
\usepackage{amsmath}
\usepackage{color}



\begin{document}

\newcommand{\be}{\begin{equation}}
\newcommand{\ee}{  \end{equation}}
\newcommand{\ba}{\begin{eqnarray}}
\newcommand{\ea}{  \end{eqnarray}}
\newcommand{\ve}{\varepsilon}

\title{Spin transport through a nanojunction with a precessing anisotropic molecular spin: Quantum interference and spin-transfer torque}

\author{Milena Filipovi\'{c}}
\affiliation{Institute of Physics Belgrade, University of Belgrade, Pregrevica 118, 11080 Belgrade, Serbia}

\date{\today}

\begin{abstract}
The subject of this study is spin transport through a molecular orbital connected to two leads, and coupled via exchange interaction with a precessing anisotropic molecular spin in a constant magnetic field. 
The inelastic spin-flip processes between molecular quasienergy levels are driven by the molecular spin precession. By setting the Larmor frequency, the tilt angle of molecular magnetization with respect to the magnetic field, and the magnetic anisotropy parameter, one can modulate the spin current and noise, spin-transfer torque, and related torque coefficients. Moreover, the dc-spin current and spin-transfer torque components provide the quasienergy level structure in the orbital. Quantum interference effects between states connected with spin-flip processes manifest themselves as dips (minimums) and peaks (maximums) in spin-current noise, matching Fano-like resonance profiles with equal probabilities of interfering elastic and inelastic spin-flip pathways. By proper adjustment of the anisotropy parameter and magnetic field, the precession is suppressed and the torque vanishes, revealing the anisotropy parameter via a dc-spin current or torque measurement. The results of the study show that spin transport and spin-transfer torque can be manipulated by the anisotropy parameter even in the absence of the magnetic field. 
\end{abstract}

\maketitle

\section{Introduction}

Due to small size and uniaxial magnetic anisotropy which leads to magnetic bistability, single-molecule magnets are potential candidates for magnetic storage and information processing.\cite{m1,m2,m3,m4,m5,m6,new1,coulombblockade2,ele1,t1,new5,new6,new7,new8,new9} Since energy barrier to molecular spin reversal depends on the magnetic anisotropy parameter,\cite{m2,m3,m7,m8} it is important to find ways to control it, e.g., via charge current or electric field.\cite{coulombblockade2,ele1,t1,an1,an2,an3,an4,ele2,ele3,ele4} For the potential applications in spintronics, various phenomena in magnetic structures have been subject of research, such as spin relaxation,\cite{new10,new11,new12,new13,new14} spin fluctuations,\cite{new15,new16,new17} geometrical spin torque,\cite{new18,new19} self-induced torque,\cite{new191} and the dynamics of magnetization driven by external means.\cite{new20,new211,new212,new213,new214,new21,new22,new23,new24,new25,new26,new27,new28} The control of magnetization in junctions by spin-polarized current was first theoretically suggested\cite{new29,new30} and then experimentally confirmed.\cite{new31,new32} By applying spin torque, the magnetization dynamics can be manipulated,\cite{new29} and as a back action the spin pumping occurs.\cite{new33,new34} It is possible to reverse magnetization via current-induced spin-transfer torque (STT).\cite{new20,new35,new36,new37,new38,new381,new382} For instance, using spin-polarized current through a single-molecule magnet connected with ferromagnetic electrodes, its spin states can be switched.\cite{m8,new39,new40,new41,new411} It has been shown that the anisotropic molecular spin can be reversed even in the absence of a magnetic field by turning on a bias voltage for one ferromagnetic and one paramagnetic lead.\cite{new42} Spin-polarized currents exert STTs on the magnetization of magnetic nanostructures in the form of field-like or damping torques.\cite{new20,new43,new44,new45,new46,new47,new48,new49} The effect of superconductivity on the magnetization dynamics has also been studied since the beginning of the new century.\cite{new24,new50,new51,new52,new53,new54,new55,new56}
	
The nonequilibrium Green's functions (NEGF) formalism\cite{Jauho1993,Jauho1994,JauhoBook} has been employed in molecular spintronics in investigations of, e.g., spin pumps,\cite{new57,new58} quantum interference,\cite{new59} spin-flip inelastic tunneling,\cite{new60,new61,we2013,we2016} and magntic skyrmion dynamics.\cite{new62,new63,new631} The NEGF technique has also been used in the theoretical calculations of spin-current shot noise.\cite{new64,new65,we2018} While investigating transport through single-molecule magnets, many effects were analysed, such as the Kondo effect,\cite{kondo1,stretching,kondo2,kondo3,kondo4} the spin-Seebeck effect,\cite{seebeck1,seebeck2,seebeck3} STT,\cite{new69,new70} and spin blockade.\cite{new42,spinblockade2,spinblockade3}  In molecular spintronics, experimental studies of spin-polarized currents,\cite{new71,new72} spin interactions,\cite{new73,new74} spin valves,\cite{new75} and spin-flip inelastic electron tunneling spectroscopy\cite{coulombblockade2,new76} have been done.

The classical magnetization dynamics is usually described by the Landau-Lifshitz-Gilbert (LLG) equation.\cite{new761,new762,new763} Even thermal effects on the magnetization dynamics have been studied using LLG equation.\cite{new77,new78,new79} The contribution of STT due to spin-polarized currents can be included in the LLG equation,\cite{new20,new44} and has been derived for molecular magnets,\cite{new21,new79,new83,bodecurrentinduced} and other magnetic systems such as spin valves or magnetic multilayers,\cite{new29} magnetic domain walls,\cite{new80} slowly varying magnetization,\cite{new81} and magnetic skyrmions.\cite{new631,new82} In quantum transport calculations, the semiclassical approach is often used, where the local magnetization of a magnetic nanostructure is treated as classical and its dynamics is described by the LLG equation, while the spin of the conduction electrons is considered as quantum.\cite{new13,new19,bodecurrentinduced,new85,t6,new86,new87,t8} It is assumed that the spin-polarized currents are carried by electrons that are fast in comparison to the local magnetization dynamics.

The aim of this article is to theoretically study the spin transport through a single molecular orbital of a molecular magnet with a precessing anisotropic spin in a constant magnetic field. The precession frequency  of the molecular spin involves Larmor frequency and a term with the uniaxial magnetic anisotropy parameter. The spin in the orbital and the molecular spin are coupled via exchange interaction. The orbital is connected to two normal metal leads, leading to spin tunneling. An STT is then exerted onto the molecular spin by the inelastic spin currents. As a back action, the molecular spin pumps spin currents into the leads. Using external means to compensate the effect of STT on the molecular spin dynamics, the spin precession remains steady. The spin currents, noise of $z$-polarized spin current, and STT are calculated using the the Keldysh NEGF technique.\cite{Jauho1993,Jauho1994,JauhoBook} The initially single molecular orbital results in four quasienergy levels dependent on the magnetic anisotropy parameter,\cite{4rad} obtained by the Floquet theorem.\cite{Floquet1,Floquet2,Floquet3,Floquet4}
The elastic spin currents are driven by the bias voltage. The inelastic spin currents, driven by the molecular spin precession, contribute to the STT. They involve electron spin-flip events accompanied by an energy change that depends on the anisotropy parameter.
The setup can be used to generate and control spin currents and STT by adjusting the anisotropy parameter, the tilt angle of the molecular spin from the magnetic field, and the Larmor frequency. Furthermore, if the anisotropy contribution to the precession frequency coincides with the Larmor frequency, the precession is suppressed, and consequently the STT vanishes. Similarly to charge-current noise,\cite{we2018,4rad} the peaks and dips in spin-current noise, resembling Fano-like line shapes,\cite{Fano2024,interf2024} occur due to quantum interference between the states connected with spin-flip events. The spin-current and noise, STT, and torque coefficients vanish for large anisotropy parameter, and they can be controlled by the anisotropy parameter in the absence of a magnetic field as well.

The remainder of the article is organised as follows. The model setup is introduced in Sec. II. The theoretical framework based on the Keldysh NEGF technique, used to calculate expressions for spin currents, noise of $z$-polarized spin current, and STT, is presented in Sec. III. The results are discussed in Sec. IV, where the properties of the $z$-polarized spin current, the corresponding autocorrelation noise, STT, and the torque coefficients are analyzed at zero temperature. The conclusions are given in Sec. V.

\section{Model Setup}

\begin{figure}
\includegraphics[width=8.5cm,keepaspectratio=true]{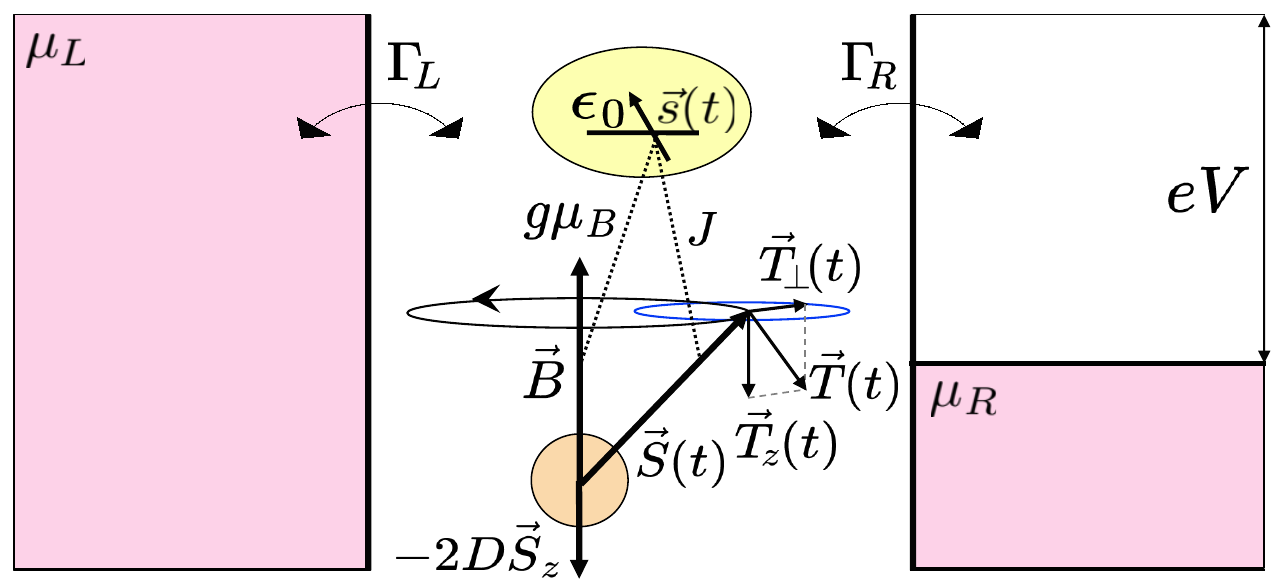}
\caption{Spin tunneling through a single molecular orbital with energy $\epsilon_{0}$, coupled to the molecular spin $\vec{S}(t)$ with anisotropy parameter $D$, via  exchange interaction with the coupling constant $J$, in the presence of a magnetic field $\vec{B}$, connected to two leads with chemical potentials $\mu_{L}$ and $\mu_R$, $eV=\mu_{L}-\mu_R$, with tunnel rates $\Gamma_L$ and $\Gamma_R$. The molecular spin $\vec{S}(t)$ precesses around the magnetic field axis with frequency $\omega=\omega_L-2DS_z$. The spin-transfer torque $\vec{T}(t)$ is exerted on the spin $\vec{S}(t)$ by the spin currents from the leads.}\label{fig: system}
\end{figure}

The junction consists of a single orbital of an anisotropic magnetic molecule in a magnetic field, connected to two noninteracting metallic leads (see Fig.~1). The magnetic field is constant, directed along the $z$-axis, $\vec{B}=B\vec{e}_{z}$, and does not affect the leads (left and right) with chemical potentials $\mu_{\xi}$, $\xi=L,R$. The system Hamiltonian is given by $\hat{H}=\hat{H}_{L}+\hat{H}_{R}+\hat{H}_{T}+\hat{H}_{\rm MO}+\hat{H}_{S}$. The first two terms represent Hamiltonians of the leads, $\hat{H}_\xi=\sum_{k,\sigma}\epsilon_{k\xi} \hat{c}^\dagger_{k\sigma\xi} \hat{c}_{k\sigma\xi}$, with $\sigma=\uparrow,\downarrow=1,2=\pm 1$ denoting the electron spin state (up or down). The third term in the Hamiltonian, $\hat{H}_{T}$, represents the tunnel coupling between the orbital of the molecule and the leads, and it can be written as  $\hat{H}_{T}=\sum_{k,\sigma,\xi}  [V_{k\xi}\hat{c}^\dagger_{k\sigma\xi} \hat{d}_{\sigma}+V^{\ast}_{k\xi}\hat{d}^\dagger_{\sigma} \hat{c}_{k\sigma\xi}]$, with matrix element $V_{k\xi}$, and
creation (annihilation) operators of the electrons in the leads and orbital $ \hat{c}^\dagger_{k\sigma\xi}(\hat{c}_{k\sigma\xi})$ and $ \hat{d}^\dagger_{\sigma} (\hat{d}_{\sigma})$. The Hamiltonian of the molecular orbital $\hat{H}_{\rm MO}$ consists of three terms, one representing the noninteracting orbital with energy $\epsilon_0$, the second representing the spin of the electron in the orbital in the presence of the magnetic field, and the third term representing the exchange interaction between the spin of the electron in the orbital and molecular spin, with the exchange coupling constant $J$, $\hat{H}_{\rm MO}=\sum_{{\sigma}}\epsilon_{0}\hat{d}^\dagger_{\sigma} \hat{d}_{\sigma}+(g\mu_{B}/\hbar)\hat{\vec{s}}\vec {B}+J\hat{\vec{s}}\vec{S}$. The spin of the electron in the molecular orbital is given by $\hat{\vec{s}}=(\hbar/2)\sum_{\sigma\sigma'}(\hat{\vec\sigma})_{\sigma\sigma'}\hat d^\dagger_\sigma\hat d_{\sigma'}$, with the vector of the Pauli matrices $\hat{\vec\sigma}=(\hat{\sigma}_x,\hat{\sigma}_y,\hat{\sigma}_z)^T$. The constants $g$ and $\mu_{B}$ are the gyromagnetic ratio of the electron, which is assumed to be equal to that of the molecular spin, and the Bohr magneton. The last term in the model Hamiltonian describes the Hamiltonian of the anisotropic spin of the molecule, $\hat{H}_{S}=(g\mu_{B}/\hbar){\hat{\vec S}}{\vec B}-D\hat{S^2_z}$, with molecular spin operator $\hat{\vec{S}}=\hat{S_x}\vec{e}_x+\hat{S_y}\vec{e}_y+\hat{S_z}\vec{e}_z$. Here, the uniaxial magnetic anisotropy parameter is given by $D$, and $\vec{e}_j$ is the unit vector along the axis $j$, with $j=x,y,z$.

The spin of the magnetic molecule is large and regarded as a classical variable $\vec{S}$, with constant length $S=\textbar{\vec{S}}\textbar\gg\hbar$, neglecting the quantum fluctuations. The vector $\vec{S}$ is the expectation value of the previously mentioned molecular spin operator, $\vec{S}=\langle\hat{\vec{S}}\rangle$, with the dynamics expressed by the Heisenberg equation of motion $\dot{\vec{S}} =\big \langle \dot{\hat{\vec{S}}} \big \rangle\ = ( i/\hbar)\big \langle \big [\hat{H},\hat{\vec{S}} \big ] \big \rangle$. To keep the molecular spin dynamics unaffected by the loss of the magnetic energy due to the exchange interaction with the spin of the tunneling electrons, one needs to use external means, e.g., radiofrequency fields.\cite{Kittel} Taking into account the Larmor precession frequency around the magnetic field axis, $\omega_{L}=(g\mu_{B}/\hbar)B$, the equation $\dot{\vec{S}}=(g\mu_{B}/\hbar)\vec{B}\times\vec{S}-2D\vec{S}_{z}\times\vec{S}$ is obtained,\cite{bodecurrentinduced} showing that the molecular spin precesses around $z$-axis with frequency $\omega=\omega_L-2DS_z$. 
The dynamics of the molecular spin can be expressed as $\vec S(t)=S_{\bot}\cos (\omega t)\vec e_x+S_{\bot}\sin (\omega t)\vec e_y+S_{z}\vec e_z$, with $S_{\bot}=S\sin\theta$ and $S_z=S\cos\theta$, where $\theta$ is the tilt angle between the positive  direction along $z$-axis and $\vec{S}$. While the precession of the molecular spin is kept undamped, i.e., the exerted STT by the flow of electron spins from the leads is externally compensated,\cite{Kittel} the precessing spin affects spin of the itinerant electrons during the exchange interaction, as a back action. Thus, the molecular spin pumps spin currents into the leads, having an impact on spin-transport properties of the junction.  

In molecular magnets realized by transition metals or rare earth elements, the Coulomb interaction in the $s$ and $p$ orbitals in the molecular ligands can be neglected.\cite{an2,new23,Godfrin} They form the highest occupied or lowest unoccupied molecular orbital, which can be considered as a localized molecular level.
The molecular spin is formed out of localized $d$ or $f$ orbitals from a transition-metal or rare-earth element, thus allowing separation between the molecular level and the localized spin.

\section{Theoretical Framework}
\subsection{Spin current}
The spin-current operators of the lead $\xi$ are given by the Heisenberg equation 
	\begin{equation}
		\hat{I}_{\xi j}(t)=\frac{\hbar}{2}\frac{d\hat{N}_{\xi j}}{dt}=\frac{i}{2}[\hat{H},\hat{N}_{\xi j}],\label{eq: commutator}
	\end{equation}
where $j=x,y,z$ denotes the component of the spin current with electronic spins oriented along the given spatial direction, $[\ , \ ]$ symbolizes the commutator, whereas $\hat{N}_{\xi j}=\sum_{{k,\sigma,\sigma\prime}}
\hat{c}^\dagger_{k\sigma\xi}(\hat{\sigma}_j)_{\sigma\sigma\prime}\hat{c}_{k\sigma\xi}$ denotes the spin occupation number operator of the lead $\xi$, with matrix elements of the Pauli oparators $(\hat{\sigma}_j)_{\sigma\sigma\prime}$. The average spin current, as a flow of spins oriented along the $j$ direction from the contact $\xi$ to the molecular orbital, can be written as
\begin{equation}
	I_{\xi j}(t) =\frac{1}{2}\bigg \langle \frac{d}{dt} \hat{N}_{\xi j}\bigg \rangle = \frac{i}{2} 
	\big \langle \big [\hat{H},\hat{N}_{\xi j} \big ] \big \rangle,
\end{equation}
in units in which $\hbar=e=1$. Employing the Keldysh NEGF technique,\cite{Jauho1994, JauhoBook} the components of the spin current can be calculated as 
\begin{align}
	\label{eq: general current}
	I_{\xi j}{(t)} =&-{\rm Re}\int dt^\prime {\rm Tr}\big\{ \hat{\sigma}_{j}[{\hat{G}}^{r} {(t,t^\prime)}{\hat{\Sigma}}^{<}_{\xi}(t^\prime,t)\nonumber\\
	& \quad \quad \quad\quad \quad\quad +{\hat{G}}^{<} {(t,t^\prime)}{\hat{\Sigma}}^{a}_{\xi}(t^\prime,t)] \big\}.
\end{align}
Here, $\hat{G}^{r,a,<,>}(t,t^\prime)$ denote the retarded, advanced, lesser, and greater Green's functions of the spin carriers in the molecular orbital. The matrix elements of the Green's functions are given by 
$G^{r,a}_{\sigma\sigma^\prime} {(t,t^\prime)}=\mp i\theta(\pm t \mp t^\prime)\langle\{\hat{d}_{\sigma}{(t)},
\hat{d}^\dagger_{\sigma^\prime} {(t^\prime)}\}\rangle$,  $G^<_{\sigma\sigma^\prime} (t,t^\prime)= i \langle \hat{d}^\dagger_{\sigma^\prime} (t^\prime) \hat{d}_\sigma(t)\rangle$, and
$G^>_{\sigma\sigma^\prime} (t,t^\prime)= -i \langle \hat{d}_{\sigma} (t) \hat{d}^\dagger_{\sigma^\prime}(t^\prime)\rangle$, where $\{\cdot ,\cdot\}$ symbolizes the anticommutator. 
The self-energies from the tunnel coupling between the orbital 
and lead $\xi$ are represented by $\hat{\Sigma}^{r,a,<,>}_{\xi}(t,t^\prime)$, with diagonal matrix elements in the electron spin space with respect to the basis of the eigenstates of $\hat{s}_{z}$. Their nonzero matrix elements can be expressed as $\Sigma^{r,a,<}_{\xi}(t,t')=\sum_{{k}}
V_{k\xi}^{\phantom{\ast}}g^{r,a,<,>}_{k\xi}{(t,t')}V^{\ast}_{k\xi}$, with $g^{r,a,<,>}_{k \xi}{(t,t')}$ denoting the Green's functions of the electrons in the lead $\xi$.
Applying the double Fourier transformations in Eq.~(3), it can be further simplified as
\begin{align}\label{eq: struja}
	I_{\xi j}{(t)}=&\,\Gamma_{\xi}{\rm Im}\int\frac{d\epsilon}{2\pi}\int\frac{d\epsilon^\prime}{2\pi} e^{-i(\epsilon-\epsilon^\prime)t}\nonumber\\
	&\times{\rm Tr}\bigg\{\hat{\sigma}_{j}\bigg [f_{\xi}(\epsilon^{\prime}){\hat{G}}^{r} {(\epsilon,\epsilon^{\prime})+ {\frac{1}{2}{\hat{G}}^{<}} {(\epsilon,\epsilon^{\prime})}}\bigg ]\bigg\},
\end{align}
with the tunnel coupling between the orbital and lead $\xi$, $\Gamma_{\xi}(\epsilon)=2\pi\sum_{{k}}\lvert V_{k\xi}\rvert^{2}\delta (\epsilon-\epsilon_{k\xi})$, which is energy independent and constant in the wide-band limit. The Fermi-Dirac distribution of the spin carriers in the lead $\xi$ is given by $f_{\xi}(\epsilon)=[e^{(\epsilon-\mu_{\xi})/k_{B}T}+1]^{-1}$,
where $T$ and $k_{B}$ are the temperature and the Boltzmann constant.

The retarded Green's function of the electrons in the orbital of the molecule can be calculated applying Dyson's expansion and analytic continuation rules.\cite{JauhoBook} Its double-Fourier transformed matrix elements can be expressed as\cite{Guo,we2016,4rad}
	\begin{align}\label{eq: greenn}
		G^{r}_{\sigma\sigma}(\epsilon,\epsilon^\prime)&
		=\frac{2\pi \delta(\epsilon-\epsilon^\prime)G^{0r}_{\sigma\sigma}(\epsilon)}{1-\gamma^{2}G^{0r}_{\sigma\sigma}(\epsilon)
			{G^{0r}_{-\sigma-\sigma}(\epsilon_{\sigma})}},\\
		\label{eq: greenn1}	G^{r}_{\sigma-\sigma}(\epsilon,\epsilon^\prime)&
		=\frac{2\pi\gamma\delta(\epsilon_{\sigma}-\epsilon^{\prime})G^{0r}_{\sigma\sigma}(\epsilon)
			G^{0r}_{-\sigma-\sigma}(\epsilon_{\sigma})}{
			1-\gamma^{2}G^{0r}_{\sigma\sigma}(\epsilon)
			G^{0r}_{-\sigma-\sigma}(\epsilon_{\sigma})},
	\end{align}
with $\epsilon_{\sigma}=\epsilon-\sigma\omega=\epsilon-\sigma (\omega_L-2DS_{z})$, $\gamma=JS\sin(\theta)/2$. The Fourier transformed retarded Green's function of the electrons in the molecular orbital in the presence of the static molecular spin, $S=S_z$, calculated using the equation of motion technique,\cite{Bruus} is given by $\hat{G}^{0r}(\epsilon)=[\epsilon-\epsilon_{0}-\Sigma^{r}-\hat{\sigma}_{z} (g\mu_{B}B+J S_{z})/2]^{-1}$,\cite{Guo,bodecurrentinduced}
 where $\Sigma^{r,a}=\mp i\Gamma/2$ and $\Gamma=\sum_{\xi}\Gamma_{\xi}$.
The Green's functions $\hat{G}^{<,>}(\epsilon,\epsilon^\prime)$ can be obtained using the double-Fourier transformed Keldysh equation, expressed as
$\hat{G}^{<,>}(\epsilon,\epsilon^\prime)=\int d\epsilon^{\prime\prime}\hat{G}^{r}(\epsilon,\epsilon^{\prime\prime})\hat{\Sigma}^{<,>}(\epsilon^{\prime\prime})\hat{G}^{a}(\epsilon^{\prime\prime},\epsilon^\prime)/2\pi$,\cite{JauhoBook} with lesser self-energy ${\Sigma}^{<}(\epsilon)=i\sum_{\xi}\Gamma_{\xi}f_{\xi}(\epsilon)$, greater self-energy ${\Sigma}^{>}(\epsilon)=i\sum_{\xi}\Gamma_{\xi}(f_{\xi}(\epsilon)-1)$ and advanced Green's function $\hat{G}^{a}(\epsilon,\epsilon^\prime)=[\hat{G}^{r}(\epsilon^\prime,\epsilon)]^\dagger$.

The spin current given by Eq.~(4) can be calculated using the above expressions for the Green's functions $\hat{G}^{r}(\epsilon,\epsilon^\prime)$ and $\hat{G}^{<}(\epsilon,\epsilon^\prime)$. As a result, the time-dependent spin-current components $I_{\xi x}(t)$ and $I_{\xi y}(t)$ read
\begin{align}
	\label{eq: x-complex} I_{\xi x}(t)=&I_{\xi x}(D)e^{-i(\omega_{L}-2DS_{z}) t}+I^*_{\xi x}(D)e^{i(\omega_{L}-2DS_{z}) t},\\
	\label{eq: y-complex} I_{\xi y}(t)=&I_{\xi y}(D)e^{-i(\omega_{L}-2DS_{z}) t}+I^*_{\xi y}(D)e^{i(\omega_{L}-2DS_{z}) t},
\end{align}	
while $I_{\xi z}$ is time-indnpendent. The expressions for $I_{\xi z}$ and the complex functions $I_{\xi x}(D)$ and $I_{\xi y}(D)$ are presented by Eqs.~(A1)--(A3) in the Appendix. For the isotropic molecular spin ($D=0$), they reduce to the expressions obtained before.\cite{we2016}

In the presence of the precessing anisotropic molecular spin $\vec{S}(t)$ and the external magnetic field $\vec{B}$, the initial single resonant transmission channel with energy $\epsilon_0$ results in four channels available for spin transport. They are located at Floquet quasienergies,\cite{4rad} 
	\begin{align}
		\epsilon_{1,3}=\epsilon_{0}-\frac{\omega_{L}}{2}+&DS_{z}\pm\sqrt{D(D+J)S^{2}_{z}+\bigg (\frac{JS}{2}\bigg )^2},\\
		\epsilon_{2,4}=\epsilon_{0}+\frac{\omega_{L}}{2}-&DS_{z}\pm\sqrt{D(D+J)S^{2}_{z}+\bigg (\frac{JS}{2}\bigg )^2},
	\end{align}
	obtained using the Floquet theorem,\cite{Floquet1,Floquet2,Floquet3,Floquet4} since the Hamiltonian of the molecular orbital is a periodic function of time  $\hat{H}_{\rm MO}(t)=\hat{H}_{\rm MO}(t+2\pi/\omega)$. As a result of the exchange interaction between the spin of the molecule and the itinerant electron spin, the $\downarrow$ state with quasienergy $\epsilon_{1}(\epsilon_3)$ is coupled to the $\uparrow$ state with quasienergy $\epsilon_{2}(\epsilon_4)=\epsilon_{1}(\epsilon_3)+\omega=\epsilon_{1}(\epsilon_3)+\omega_{L}-2DS_{z}$. Namely, periodic motion of the molecular spin leads to the absorption (emission) of an energy quantum $\omega$ by the spin-carrying electron in the orbital, accompanied by a spin-flip. 

\subsection{Noise of $z$-polarized spin current}

To obtain further characteristics of spin transport, one can study spin-current noise. The noise of spin current polarized along the $z$-direction is calculated here, as a complement to previously studied charge-current noise.\cite{4rad} 
Since only the tunneling Hamiltonian $\hat{H}_T$ contributes to the commutator in Eq.~(1), the resulting spin-current operator $\hat{I}_{\xi z}(t)$ is given by 
\begin{equation}
	\hat{I}_{\xi z}(t)=\frac{i}{2}\sum_{\sigma}(-1)^{\sigma}\hat{I}_{\xi\sigma}(t),\label{eq: sigmacomponents}
\end{equation}
where the operator component $\hat{I}_{\xi\sigma}(t)$ reads
\begin{equation}
	\hat{I}_{\xi\sigma}(t)=\sum_{k}[V_{k\xi}\hat{c}^{\dag}_{k\sigma\xi}(t)\hat{d}_{\sigma}(t)-V^{*}_{k\xi}\hat{d}^{\dag}_{\sigma}(t)\hat{c}_{k\sigma\xi}(t)].
\end{equation}
The spin-current fluctuation operator $\delta\hat{I}_{\xi z}(t)$ in lead $\xi$ can be written as
\begin{equation}
	\delta\hat{I}_{\xi z}(t)=\hat{I}_{\xi z}(t)-\langle\hat{I}_{\xi z}(t)\rangle.\label{eq: correlation}
\end{equation}
The nonsymmetrized noise of $z$-polarized spin current, defined as a correlation between fluctuations of $z$-polarized spin currents in contacts ${\xi}$ and ${\zeta}$, is given by\cite{new64,Sauret}
\begin{equation}
	S^{zz}_{\xi\zeta}(t,t')=\langle\delta\hat{I}_{\xi z}(t)\delta\hat{I}_{\zeta z}(t')\rangle,\label{eq: 5.4}
\end{equation}	
whereas the symmetrized noise can be written as
\begin{equation}
	S^{zz}_{\xi\zeta S}(t,t')=\frac{1}{2}\langle\{\delta\hat{I}_{\xi z}(t),\delta\hat{I}_{\zeta z}(t')\}\rangle.\label{eq: symmetrized_noise}
\end{equation}	
According to Eqs.~(11) and (13), the nonsymmetrized noise of $z$-polarized spin current equals\cite{new64,Sauret} 
\begin{equation}
	S^{zz}_{\xi\zeta}(t,t')=-\sum_{\sigma\sigma'}(-1)^{\delta_{\sigma\sigma'}}
	S^{\sigma\sigma'}_{\xi\zeta}(t,t'),\label{eq: 5.6}
\end{equation}
with $S^{\sigma\sigma'}_{\xi\zeta}(t,t')=(-1/4)\langle\delta\hat{I}_{\xi\sigma}(t)\delta\hat{I}_{\zeta\sigma'}(t')\rangle$.
Implementing Wick's theorem\cite{Wick2024} and Langreth analytical continuation rules\cite{Langreth2024} to the correlation function $S^{\sigma\sigma'}_{\xi\zeta}(t,t')$ in Eq.~(16), and using the Green's functions of the molecular orbital and the self-energies from the tunnel couplings between the orbital and the leads, one obtains the 
expression for the noise of $z$-polarized spin current.\cite{we2018}
Employing the Fourier transforms of the Green's functions $G^{r,a,<,>}_{\sigma\sigma'}(\epsilon,\epsilon')$ and self-energies  $\Sigma^{r,a,<,>}_{\xi}(\epsilon)$, this expression can be transformed into 
\begin{widetext}
\begin{align}
	S^{zz}_{\xi\zeta}(t,t')=\frac{1}{4}\sum_{\sigma\sigma'}(-1)^{\delta_{\sigma\sigma'}}\bigg\{\int&\frac{d\epsilon_{1}}{2\pi}\int \frac{d\epsilon_{2}}{2\pi}\int \frac{d\epsilon_{3}}{2\pi}	\int \frac{d\epsilon_{4}}{2\pi}e^{-i(\epsilon_{1}-\epsilon_{2})t}e^{i(\epsilon_{3}-\epsilon_{4})t'}\nonumber\\
	\times&\big\{[G^{r}_{\sigma\sigma'}(\epsilon_{1},\epsilon_{3})\Sigma^{>}_{\zeta}(\epsilon_{3})+2G^{>}_{\sigma\sigma'}(\epsilon_{1},\epsilon_{3})\Sigma^{a}_{\zeta}]
	[G^{r}_{\sigma'\sigma}(\epsilon_{4},\epsilon_{2})\Sigma^{<}_{\xi}(\epsilon_{2})+2G^{<}_{\sigma'\sigma}(\epsilon_{4},\epsilon_{2})\Sigma^{a}_{\xi}]\nonumber\\
	&+[\Sigma^{>}_{\xi}(\epsilon_{1})G^{a}_{\sigma\sigma'}(\epsilon_{1},\epsilon_{3})+2G^{>}_{\sigma\sigma'}(\epsilon_{1},\epsilon_{3})\Sigma^{r}_{\xi}]
	[\Sigma^{<}_{\zeta}(\epsilon_{4})G^{a}_{\sigma'\sigma}(\epsilon_{4},\epsilon_{2})+2G^{<}_{\sigma'\sigma}(\epsilon_{4},\epsilon_{2})\Sigma^{r}_{\zeta}]\nonumber\\
	&+4\Sigma^{r}_{\xi}\Sigma^{a}_{\zeta}G^{>}_{\sigma\sigma'}(\epsilon_{1},\epsilon_{3})G^{<}_{\sigma'\sigma}(\epsilon_{4},\epsilon_{2})\big\}\nonumber\\
	&-\delta_{\xi\zeta}\delta_{\sigma\sigma'}\int \frac{d\epsilon_{1}}{2\pi}\int \frac{d\epsilon_{2}}{2\pi}\int \frac{d\epsilon_{3}}{2\pi}\nonumber\\
	&\hspace{0.4cm}\times\big\{e^{-i(\epsilon_{1}-\epsilon_{3})t}e^{i(\epsilon_{2}-\epsilon_{3})t'}G^{>}_{\sigma\sigma'}(\epsilon_{1},\epsilon_{2})\Sigma^{<}_{\xi}(\epsilon_{3})\nonumber\\
	&\hspace{0.4cm}+e^{-i(\epsilon_{1}-\epsilon_{3})t}e^{i(\epsilon_{1}-\epsilon_{2})t'}\Sigma^{>}_{\xi}(\epsilon_{1})G^{<}_{\sigma'\sigma}(\epsilon_{2},\epsilon_{3})\big\}\bigg\}.\label{eq: neznam26maj}
\end{align}	
\end{widetext}
Since the noise $S^{zz}_{\xi\zeta}(t,t')$ depends only on the time difference $\tau =t-t'$, its power spectrum equals
\begin{equation}
	S^{zz}_{\xi\zeta}(\Omega)=\int d\tau e^{i\Omega\tau}S^{zz}_{\xi\zeta}(\tau).\label{eq: 5.17}
\end{equation}
The symmetrized noise spectrum is given by
\begin{equation}
	S^{zz}_{\xi\zeta S}(\Omega)=\frac{1}{2}[S^{zz}_{\xi\zeta}(\Omega)+S^{zz}_{\zeta\xi}(-\Omega)],\label{eq: 5.177}
\end{equation}
with $S^{\sigma\sigma'}_{\xi\zeta S}(\Omega)=\frac{1}{2}[S^{\sigma\sigma'}_{\xi\zeta}(\Omega)+S^{\sigma'\sigma}_{\zeta\xi}(-\Omega)]$. As it is of experimental interest, the zero-frequency noise power of $z$-polarized spin-current $S^{zz}_{LL}=S^{zz}_{LL}(0)=S^{zz}_{LLS}(0)$ at zero temperature will be analyzed and discussed.

\subsection{Spin-transfer torque}

In the presence of the anisotropic molecular spin $\vec{S}(t)$, interacting with the incoming flow of electron spins from the leads via exchange interactions, the transfer of spin angular momentum to the molecular spin occurs, resulting in the STT $\vec{T}(t)$ exerted on the molecular spin. As already mentioned, the STT is compensated by external means, so that the molecular spin precession remains unaffected. Considering that the total spin angular momentum is conserved, the spin of the molecule generates a torque $-\vec{T}(t)$ acting on the spin currents from the leads,\cite{new20,new29,new30,new44} 
	\begin{eqnarray} 
		-\vec{T}(t)=\vec{I}_L(t)+\vec{I}_R(t),\label{eq: torque}
		\end{eqnarray}
where $\vec{I_\xi}(t)=\sum_{{j}} I_{\xi j}(t)\vec{e}_j$, while $\vec{T}(t)=\sum_{{j}}T_{j}\vec{e}_j$. Employing Eqs.~(7), (8), (20), and Eqs.~(A1)--(A3) given in the Appendix, the spatial components of the STT, $T_j$, can be expressed as
\begin{widetext}
\begin{align}
T_{x}(t)=\label{eq: torque Tx}-&\int\frac{d\epsilon}{2\pi}\sum_{\xi\zeta}\frac{\Gamma_{\xi}\Gamma_{\zeta}}{\Gamma}[f_{\xi}(\epsilon-\omega_L+2DS_{z})-f_{\zeta}(\epsilon)]
	\times{\rm Im}\bigg\{\frac{\gamma G^{0r}_{11}(\epsilon)G^{0a}_{22}(\epsilon-\omega_L+2DS_{z})}
	{\lvert 1-\gamma^{2}G^{0r}_{11}(\epsilon)G^{0r}_{22}(\epsilon-\omega_{L}+2DS_{z})\lvert^{2}}\nonumber\\
	\times &[1-\gamma^{2}G^{0a}_{11}(\epsilon)G^{0r}_{22}(\epsilon-\omega_L+2DS_{z})]e^{-i(\omega_{L}-2DS_{z}) t}\bigg\},\\
	T_{y}(t)=\label{eq: torque Ty}-&\int\frac{d\epsilon}{2\pi}\sum_{\xi\zeta}\frac{\Gamma_{\xi}\Gamma_{\zeta}}{\Gamma}[f_{\xi}(\epsilon-\omega_L+2DS_{z})-f_{\zeta}(\epsilon)]
	\times{\rm Re}\bigg\{\frac{\gamma G^{0r}_{11}(\epsilon)G^{0a}_{22}(\epsilon-\omega_L+2DS_{z})}
	{\lvert 1-\gamma^{2}G^{0r}_{11}(\epsilon)G^{0r}_{22}(\epsilon-\omega_{L}+2DS_{z})\lvert^{2}}\nonumber\\
	\times &[1-\gamma^{2}G^{0a}_{11}(\epsilon)G^{0r}_{22}(\epsilon-\omega_L+2DS_{z})]e^{-i(\omega_{L}-2DS_{z}) t}\bigg\},\\
	T_{z}=\label{eq: torque Tz}-&\int\frac{d\epsilon}{2\pi}\sum_{\xi\zeta}\Gamma_{\xi}
	\Gamma_{\zeta}[f_{\xi}(\epsilon-\omega_L+2DS_{z})-f_{\zeta}(\epsilon)]\times\frac{\gamma^{2}\lvert G^{0r}_{11}(\epsilon) G^{0r}_{22}(\epsilon-\omega_L+2DS_{z})\lvert^2}
	{\lvert 1-\gamma^{2}G^{0r}_{11}(\epsilon)G^{0r}_{22}(\epsilon-\omega_{L}+2DS_{z})\lvert^{2}}.
\end{align}
\end{widetext}
For the isotropic molecular spin with $D=0$, Eqs.~(21)--(23) reduce to the previously calculated expressions for the spatial components of the STT.\cite{we2016}  
If the magnetic field is turned off ($\omega_{L}=0$), then $\vec{T}(t)=0$ for $D=0$,\cite{we2016} whereas $\vec{T}(t)\neq 0$ for $D\neq 0$ according to Eqs.~(21)--(23).

Both $T_{x}(t)$ and $T_{y}(t)$ can be written as $T_{x}(t)=T_{x}\cos(\omega t+\phi_x)$ and $T_{y}(t)=T_{y}\cos(\omega t+\phi_y)$. The phase difference $\phi_{x}-\phi_{y}=\pi/2$, whereas $T_x$ and $T_y$ are the amplitudes. Since the amplitudes are equal, it is convenient to introduce the in-plane torque component, orthogonal to the $z$-axis, $\vec{T}_{\bot}(t)=\vec{T}_{x}(t)+\vec{T}_{y}(t)$, while
	\begin{eqnarray} 
	\vec{T}(t)=\vec{T}_{\bot}(t)+\vec{T}_{z}(t).\label{eq: torque-2}
\end{eqnarray}
The in-plane component of the STT, $\vec{T}_{\bot}(t)$, precesses around the end point of the molecular spin $\vec{S}(t)$ in the $xy$ plane (see Fig.~1), and it has a constant magnitude $\lvert \vec{T}_{\bot}(t)\rvert=T_{\bot} =T_{x}=T_{y}$.

In relation to the molecular spin vector $\vec{S}(t)$, the STT can be reformulated as
\begin{eqnarray}
	\vec{T}(t)=\frac{\alpha}{S}\dot{\vec{S}}(t)\times\vec{S}(t)+\beta\dot{\vec{S}}(t)+\eta\vec{S}(t).\label{eq: torque-Gilbert}
\end{eqnarray}
The Gilbert damping component of the torque is given by the first term in Eq.~(25) , with $\alpha$ the Gilbert damping coefficient. It tends to align the anisotropic molecular spin $\vec{S}(t)$ anti(parallel) to the direction of the effective magnetic field $\vec{B}_{\rm eff}=\vec{B}-(2D/g\mu_{B})\vec{S}_z$ and dissipate(add) magnetic energy. The second term with coefficient $\beta$ tends to change the precession frequency of the molecular spin $\vec{S}(t)$ and the direction of the molecular spin precession for $\beta<0$. The contribution in the third therm is characterized by the coefficient $\eta$. The Cartesian representation of the STT given by Eq.~(25) results in spatial components
	\begin{align}
	T_{x}(t)&=\label{eq: torque Tx1}S_{\bot}\bigg(\frac{\alpha}{S}\omega S_{z}+\eta\bigg)\cos (\omega t)-\beta\omega S_{\bot}\sin (\omega t),\\
	T_{y}(t)&=\label{eq: torque Ty1}S_{\bot}\bigg(\frac{\alpha}{S}\omega S_{z}+\eta\bigg)\sin (\omega t)+\beta\omega S_{\bot}\cos (\omega t),\\
	T_{z}&=\label{eq: torque Tz1}-\frac{\alpha}{S}\omega S^{2}_{\bot}+\eta S_z,
\end{align}
where $\dot{\vec{S}}(t)=\omega S_{\bot}[-\sin (\omega t)\vec{e}_{x}+\cos  (\omega t)\vec{e}_{y}]$ for the precessing molecular spin.
Combining expressions given by Eqs.~(21)--(23) and (26)--(28), the resulting torque coefficients $\alpha$ and $\beta$ can be written as 
\begin{figure*}
	\includegraphics[height=5.75cm,keepaspectratio=true]{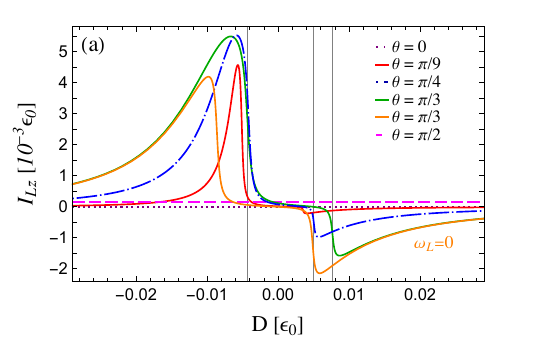}  
	\includegraphics[height=5.75cm,keepaspectratio=true]{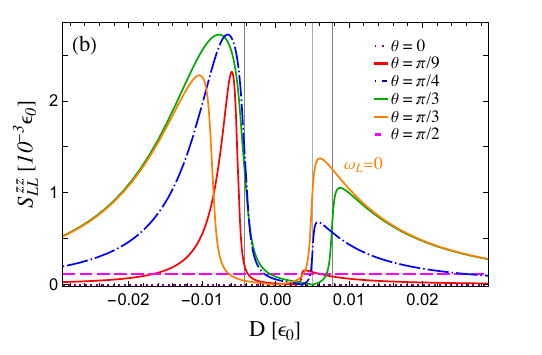}    
	\caption{(a) Spin current $I_{Lz}$ and (b) auto-correlation spin-current shot noise $S^{zz}_{LL}$, as functions of the uniaxial magnetic anisotropy parameter $D$ for different tilt angles $\theta$, at zero temperature. The magnetic field $\vec{B}=B\vec{e}_z$ and Larmor frequency $\omega_{L}=\nobreak0.5\,\epsilon_0$, except for a zero magnetic field where $\omega_{L}=0$ (orange line). The chemical potentials of the leads are equal: $\mu_{L}=\mu_{R}=\nobreak0.1\,\epsilon_0$. The other parameters are set to $\Gamma=\nobreak0.05\, \epsilon_{0},\,\Gamma_{L}=\Gamma_{R}=\Gamma/2,\, J=\nobreak0.01\,\epsilon_{0}$, and $S=100$. Grid lines for $\theta=\pi/3$ and $\omega_{L}=\nobreak0.5\,\epsilon_0$ (green line) are positioned at $D=-0.00431\,\epsilon_0$ ($\mu_{L}=\mu_{R}=\epsilon_3$), $D=0.00766\,\epsilon_0$ ($\mu_{L}=\mu_{R}=\epsilon_4$), and $D=0.005\,\epsilon_0$ ($\omega=0$).}\label{fig: spin-current-D}
\end{figure*}
\begin{widetext}
	\begin{align}
		\alpha=\label{eq: alpha}-&\frac{1}{(\omega_{L}-2DS_{z})S}\int\frac{d\epsilon}{2\pi}\sum_{\xi\zeta}\Gamma_{\xi}
		\Gamma_{\zeta}[f_{\xi}(\epsilon-\omega_L+2DS_{z})-f_{\zeta}(\epsilon)]\nonumber\\
		&\times\frac{( JS_{z}/2\Gamma){\rm Im}\{G^{0r}_{11}(\epsilon)G^{0a}_{22}(\epsilon-\omega_L+2DS_{z})\}-\gamma^{2}\lvert G^{0r}_{11}
			(\epsilon)G^{0r}_{22}(\epsilon-\omega_L+2DS_{z})\lvert^2}
		{\lvert 1-\gamma^{2}G^{0r}_{11}(\epsilon)G^{0r}_{22}(\epsilon-\omega_{L}+2DS_{z})\lvert^{2}},\\
		\beta=\label{eq: beta}-&\frac{J}{\omega_{L}-2DS_{z}}\int\frac{d\epsilon}{4\pi}\sum_{\xi\zeta}\frac{\Gamma_{\xi}
			\Gamma_{\zeta}}{\Gamma}[f_{\xi}(\epsilon-\omega_L+2DS_{z})-f_{\zeta}(\epsilon)]\nonumber\\
		&\times\frac{{\rm Re}\{G^{0r}_{11}(\epsilon)G^{0a}_{22}(\epsilon-\omega_L+2DS_{z})\}-\gamma^{2}\lvert G^{0r}_{11}(\epsilon) G^{0r}_{22}(\epsilon-\omega_L+2DS_{z})\lvert^2}
		{\lvert 1-\gamma^{2}G^{0r}_{11}(\epsilon)G^{0r}_{22}(\epsilon-\omega_{L}+2DS_{z})\lvert^{2}},
	\end{align} 
\end{widetext}
while the coefficient $\eta$ can be expressed in terms of $T_{z}$ and Gilbert damping coefficient $\alpha$ as
\begin{eqnarray}
	\eta=\frac{T_{z}}{S_{z}}+\frac{4\gamma^{2}(\omega_{L}-2DS_{z})}{J^{2}SS_{z}}\alpha.\label{eq: eta}
\end{eqnarray}
In the limit $|D|\ll|\omega_{L}/2S_{z}|$, the effect of the uniaxial magnetic anisotropy on the STT can be neglected, and Eqs.~(29)--(31) are in agreement with the coefficients $\alpha$, $\beta$, and $\eta$ obtained for the isotropic molecular spin.\cite{we2016} Finally, the magnitude of the in-plane torque component $T_\bot$ can be expressed using $\alpha$, $\beta$ and $T_z$ as follows
\begin{eqnarray}
	T_{\bot}=\frac{2\lvert\gamma(\omega_{L}-2DS_{z})\rvert}{J}\sqrt{\beta^{2}+\frac{1}{S^{2}_z}\bigg [S\alpha+\frac{T_z}{(\omega_L-2DS_{z})}\bigg ]^2}.\label{eq: T_bot}
\end{eqnarray}
For a static molecular spin with $\omega=0$, $\dot{\vec{S}}(t)=0$, and $D=\omega_{L}/2S_z$, according to Eqs.~(21)--(23) and (31), the STT vanishes, $\vec{T}(t)=\vec{0}$, and $\eta=0$. This is expected since the exchange of spin angular momentum occurs due to the exchange interaction of electron spins with the rotational component of the molecular spin.\cite{we2013,we2016} Using Eqs.~(29) and (30), one obtains nonzero  coefficients $\alpha$ and $\beta$ in the limit $\omega\rightarrow 0$, given by Eqs.~(A4) and (A5) in the Appendix.

\begin{figure}
	\includegraphics[height=5.6cm,keepaspectratio=true]{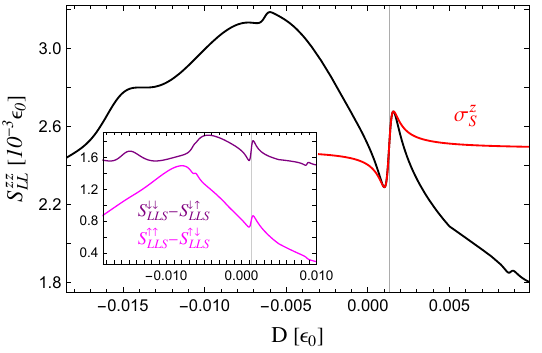}   
	\caption{Spin-current shot noise $S^{zz}_{LL}$ as a function of the magnetic anisotropy parameter $D$ at zero temperature for $\theta=\pi/3$, $\mu_{L}=0.65\,\epsilon_0$, and $\mu_{R}=0$, with $\vec{B}=B\vec{e}_z$. Around the resonant anisotropy parameter $D_{\rm res}=0.00129\,\epsilon_0$ (grid line) the spin-current noise $S^{zz}_{LL}$ (black line) matches the Fano-like shape of the resonance profile $\sigma^{z}_S$ (red line). The inset shows contributions of $S^{\uparrow\uparrow}_{LLS}-S^{\uparrow\downarrow}_{LLS}$ (pink line) and $S^{\downarrow\downarrow}_{LLS}-S^{\downarrow\uparrow}_{LLS}$ (purple line) to the resulting shape of the resonance profile in $S^{zz}_{LL}$. The other parameters are set to $\Gamma=\nobreak0.05\, \epsilon_{0},\,\Gamma_{L}=\Gamma_{R},\,\omega_{L}=0.5\,\epsilon_{0},\, J=\nobreak0.01\,\epsilon_{0}$, and $S=100$.}\label{fig: Fano_spin_resonance}
\end{figure}

\begin{figure*}
	\includegraphics[height=3.96cm,keepaspectratio=true]{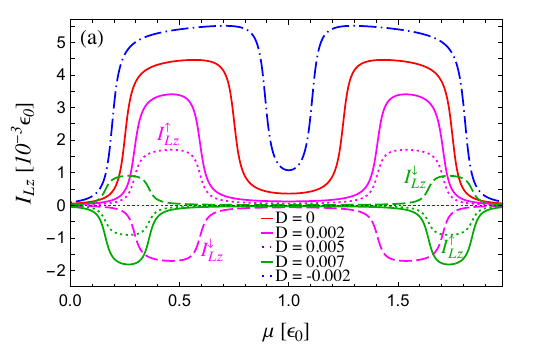}  
	\includegraphics[height=3.965cm,keepaspectratio=true]{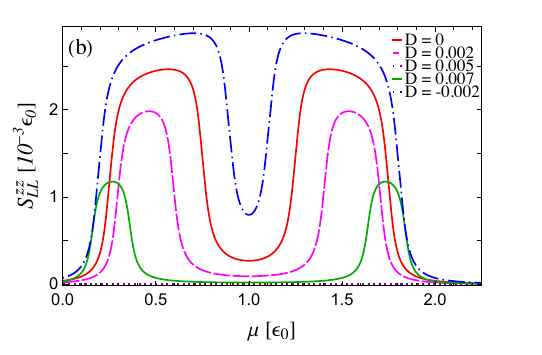} 
	\includegraphics[height=3.965cm,keepaspectratio=true]{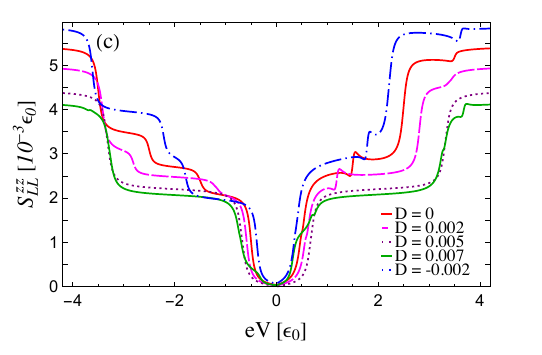}   
	\caption{(a) Spin current $I_{Lz}$, (b) auto-correlation spin-current shot noise $S^{zz}_{LL}$, as functions of the chemical potential of the leads $\mu=\mu_{L}=\mu_{R}$, and (c) auto-correlation spin-current shot noise $S^{zz}_{LL}$ as a function of the applied bias voltage $eV=\mu_{L}-\mu_{R}$, with $\mu_{L,R}=\pm eV/2$, for different uniaxial magnetic anisotropy parameters $D$, with $\vec{B}=B\vec{e}_{z}$, at zero temperature. The other parameters are set to $\Gamma=\nobreak0.05\, \epsilon_{0},\,\Gamma_{L}=\Gamma_{R}=\Gamma/2,\,\omega_L=\nobreak0.5\,\epsilon_{0},\, J=\nobreak0.01\,\epsilon_{0},\, S=100$, and $\theta=\pi/3$.  All energies are given in the units of $\epsilon_{0}$. For $\mu_{L}=\mu_{R}$ and $D=\omega_{L}/2S_z=0.005\,\epsilon_0$, $I_{Lz}=0$ and $S^{zz}_{LL}=0$ (purple dotted lines).}\label{fig: spin-current-eV}
\end{figure*}

\begin{figure*}[t]
\includegraphics[height=5.85cm,keepaspectratio=true]{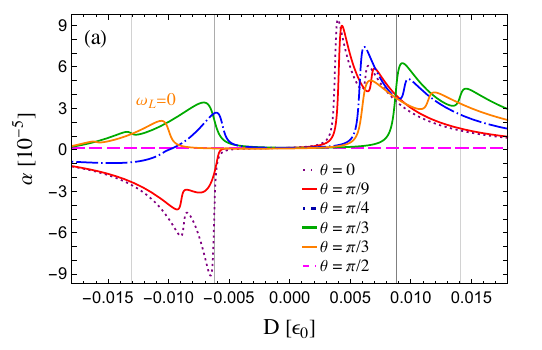}  
\includegraphics[height=5.85cm,keepaspectratio=true]{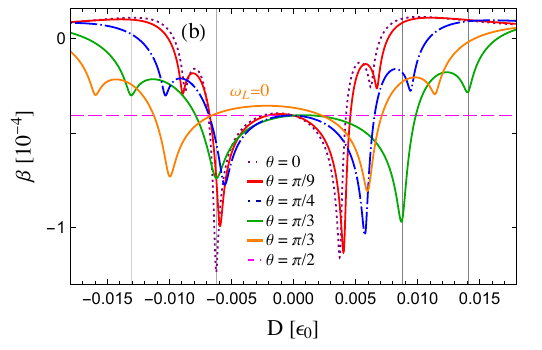} 
\includegraphics[height=5.85cm,keepaspectratio=true]{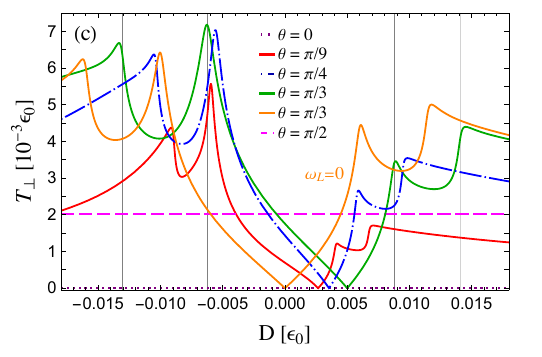} 
\includegraphics[height=5.85cm,keepaspectratio=true]{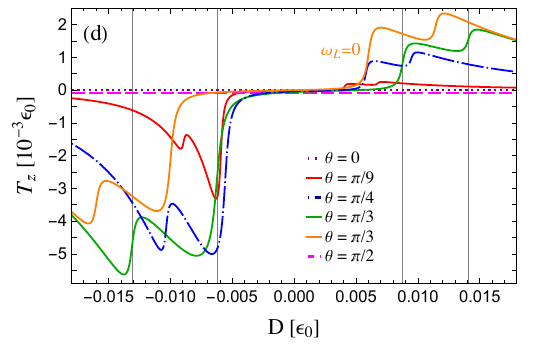}   
\caption{(a) Gilbert damping coefficient $\alpha$, (b) coefficient $\beta$, (c) magnitude of the in-plane component of the STT, $T_{\bot}$, and (d) spatial component of the torque along the $z$-direction, $T_z$, as functions of the uniaxial magnetic anisotropy parameter $D$ for different tilt angles $\theta$, at zero temperature. The magnetic field $\vec{B}=B\vec{e}_z$ and Larmor frequency $\omega_{L}=\nobreak0.5\,\epsilon_0$, except for a zero magnetic field where $\omega_{L}=0$ (orange line). The chemical potentials of the leads are equal to $\mu_{L}=\nobreak2.5\,\epsilon_0$ and $\mu_{R}=0$. The other parameters are set to $\Gamma=\nobreak0.05\, \epsilon_{0},\,\Gamma_{L}=\Gamma_{R}=\Gamma/2,\,\omega_L=\nobreak0.5\,\epsilon_{0},\, J=\nobreak0.01\,\epsilon_{0},\, S=100$. Grid lines for $\theta=\pi/3$ (green line) are positioned at $D=-0.01312\,\epsilon_0$ ($\mu_{L}=\epsilon_2$), $D=-0.00625\,\epsilon_0$ ($\mu_{R}=\epsilon_3$), $D=0.00875\,\epsilon_0$ ($\mu_{R}=\epsilon_4$), and $D=0.01406\,\epsilon_0$ ($\mu_{L}=\epsilon_1$).}\label{fig: torque-D}
\end{figure*}

\begin{figure*}
	\includegraphics[height=5.85cm,keepaspectratio=true]{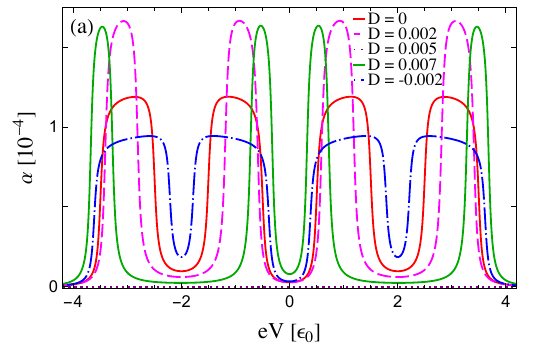}  
	\includegraphics[height=5.85cm,keepaspectratio=true]{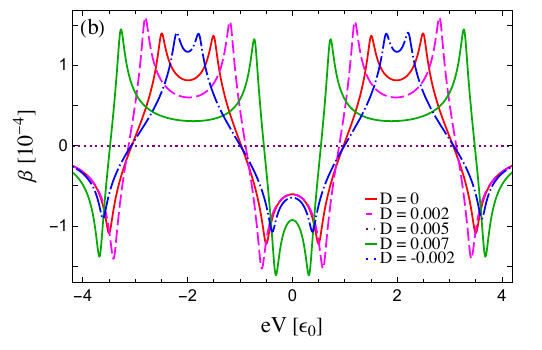} 
	\includegraphics[height=5.85cm,keepaspectratio=true]{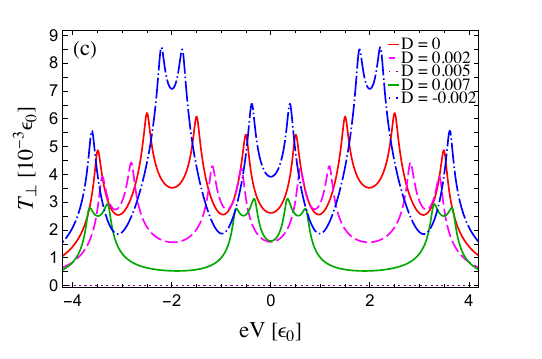} 
	\includegraphics[height=5.85cm,keepaspectratio=true]{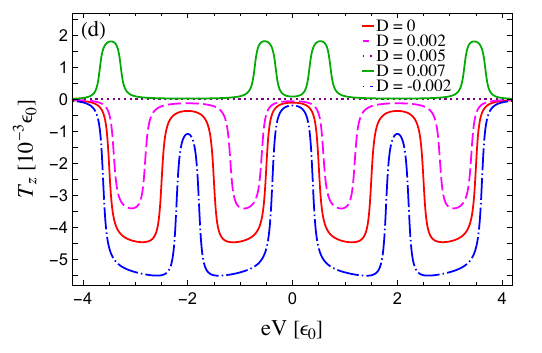}     
	\caption{(a) Gilbert damping coefficient $\alpha$, (b) coefficient $\beta$, (c) magnitude of the in-plane component of the STT, $T_{\bot}$, and (d) spatial component of the torque along the $z$-direction, $T_z$, as functions of the applied bias voltage $eV=\mu_{L}-\mu_{R}$, with $\mu_{L,R}=\pm eV/2$ and $\vec{B}=B\vec{e}_{z}$, for  different uniaxial magnetic anisotropy parameters $D$ at zero temperature. The other parameters are set to $\Gamma=\nobreak0.05\, \epsilon_{0},\,\Gamma_{L}=\Gamma_{R}=\Gamma/2,\,\omega_L=\nobreak0.5\,\epsilon_{0},\, J=\nobreak0.01\,\epsilon_{0},\, S=100$, and $\theta=\pi/3$.  All energies are given in the units of $\epsilon_{0}$. For $D=\omega_{L}/2S_z=0.005\,\epsilon_0$, $\alpha=0$, $\beta=0$, and $\vec{T}=0$ (purple dotted lines).}\label{fig: spin_eV}
\end{figure*}

\begin{figure*}
	\includegraphics[height=5.85cm,keepaspectratio=true]{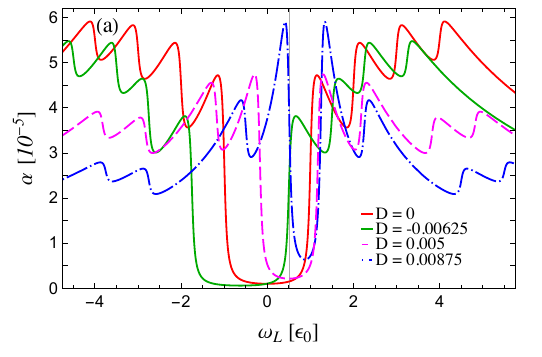}  
	\includegraphics[height=5.85cm,keepaspectratio=true]{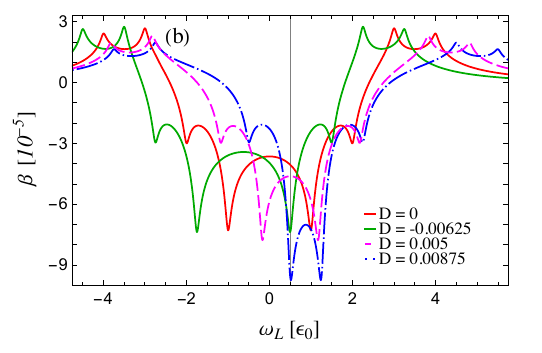} 
	\includegraphics[height=5.85cm,keepaspectratio=true]{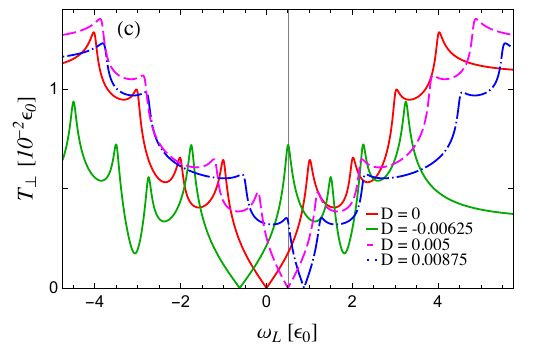} 
	\includegraphics[height=5.85cm,keepaspectratio=true]{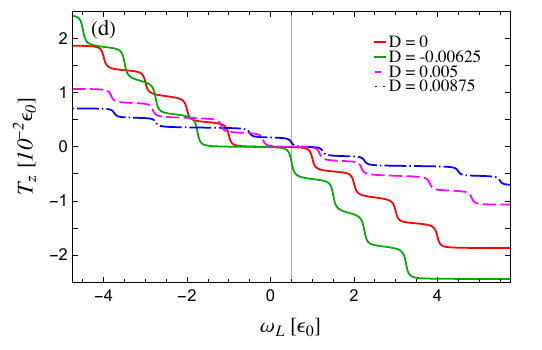}     
	\caption{(a) Gilbert damping coefficient $\alpha$, (b) coefficient $\beta$, (c) magnitude of the in-plane component of the STT, $T_{\bot}$, and (d) spatial component of the torque along the $z$-direction, $T_z$, as functions of the Larmor frequency $\omega_L$ for different uniaxial magnetic anisotropy parameters $D$. All plots are obtained at zero temperature with $\vec{B}=B\vec{e}_{z}$. The chemical potentials of the leads are equal to $\mu_{R}=0$ and $\mu_{L}=\nobreak2.5\,\epsilon_0$. The other parameters are set to $\Gamma=\nobreak0.05\, \epsilon_{0},\,\Gamma_{L}=\Gamma_{R}=\Gamma/2,\, J=\nobreak0.01\,\epsilon_{0},\, S=100$, and $\theta=\pi/3$.  All energies are given in the units of $\epsilon_{0}$. For a suppressed precession, with $\omega=0$, $\omega_{L}=2DS_z$, the STT vanishes, $T_{\bot}=0$ and $T_{z}=0$ (intersection between grid line and pink dashed line).}\label{fig: spin_Larmor}
\end{figure*} 

\section{Results and discussion}

In this section, the properties of the $z$-polarized spin current $I_{Lz}$ and autocorrelation zero-frequency noise power $S^{zz}_{LL}$ at zero temperature are discussed. Then, the characteristics of the STT generated on the anisotropic molecular spin by the spin currents from the leads, and torque coefficients, are analyzed as functions of the uniaxial magnetic anisotropy parameter $D$, bias voltage $eV=\mu_{L}-\mu_{R}$, and Larmor frequency $\omega_L$.

In Fig.~2, the spin current $I_{Lz}$ and autocorrelation noise $S^{zz}_{LL}$ are plotted as functions of the anisotropy parameter $D$, for different tilt angles $\theta$, at zero temperature and zero-bias conditions, with chemical potentials of the leads $\mu=\mu_{L}=\mu_{R}=0.1\,\epsilon_0$. 
For a static molecular spin with $\theta=0$ and $\gamma =0$ (purple dotted lines in Fig.~2), the spin current $I_{Lz}=0$ 
and $S^{zz}_{LL}=0$. In the case of $\theta=\pi/2$ (pink dashed lines in Fig.~2), the spin current $I_{Lz}$ and noise $S^{zz}_{LL}$ are independent of $D$, since $S_{z}=0$. 
For $\theta=\pi/3$ and $\omega_{L}=0.5\,\epsilon_0$ (green lines in Fig.~2), one notices maximums in $I_{Lz}$ and $S^{zz}_{LL}$ around $D=-0.00431\,\epsilon_0$, corresponding to $\mu_{L}=\mu_{R}=\epsilon_3$, while a minimum in $I_{Lz}$ and local maximum in $S^{zz}_{LL}$ around $D=0.00766\,\epsilon_0$ correspond to $\mu_{L}=\mu_{R}=\epsilon_4$ (grid lines). For $-0.00431\,\epsilon_{0}<D<0.00766\,\epsilon_{0}$, all four quasienergy levels $\epsilon_{i}$, with $i\in\{1,2,3,4\}$, lie above the chemical potentials $\mu$, so that both $I_{Lz}$ and $S^{zz}_{LL}$ drop to zero, taking into account the level broadening $\Gamma$. 
The spin current $I_{Lz}$ and noise $S^{zz}_{LL}$ vanish at $D=\omega_{L}/2S_{z}$ when $\omega=0$, as without molecular spin precession at zero-bias conditions, elastic and inelastic tunneling processes do not occur (see e.g., green line in Fig.~2 at $D=0.005\,\epsilon_0$). The spin current $I_{Lz}>0$ for $D<\omega_{L}/2S_{z}$ ($\omega>0$) since
$\downarrow$ level $\epsilon_3$ and $\uparrow$ level $\epsilon_4$ satisfy $\epsilon_{3}<\epsilon_4$, leading to positive $\uparrow$ component of spin current $I^{\uparrow}_{Lz}=-I^{\downarrow}_{Lz}>0$, and as $\mu$ approaches $\downarrow$ level $\epsilon_3$ with the increase of $D$, both $I_{Lz}$ and $S^{zz}_{LL}$ increase. Similarly, for $D>\omega_{L}/2S_{z}$ ($\omega<0$), $\downarrow$ level $\epsilon_3$ lies above $\uparrow$ level $\epsilon_4$, $\epsilon_{3}>\epsilon_{4}$ leading to $I^{\uparrow}_{Lz}=-I^{\downarrow}_{Lz}<0$, and hence $I_{Lz}<0$. In the absence of the magnetic field, $\omega_{L}=0$ (orange line in Fig.~2), and the molecular spin precesses around the $z$-axis with frequency $\omega=-2DS_{z}$. Hence, $I_{Lz}\geq 0$ for $D\leq 0$, and $I_{Lz}<0$ for $D>0$. 
After reaching resonances between $\mu$ and levels $\epsilon_3$ and $\epsilon_4$, while $\epsilon_{1}>\mu$ and $\epsilon_{2}>\mu$, both $I_{Lz}$ and $S^{zz}_{LL}$ decrease with further increase of $|D|$, as more energy is needed to flip an electron spin which is already in a lower quasienergy level, and for $|D|\gg\omega_{L}/2S_z$ they vanish. Besides, taking into account the tunneling rate $\Gamma$, for $|D|\gg\Gamma$, the precession frequency $|\omega|\gg\Gamma$, so that the probability of exchange of spin angular momentum between molecular and tunneling electron spin is low.

In Fig.~3, the spin-current shot noise $S^{zz}_{LL}$ is presented as a function of the magnetic anisotropy parameter $D$, for $\theta=\pi/3$, with chemical potentials $\mu_{L}=0.65\,\epsilon_0$ and $\mu_{R}=0$, at zero temperature. As in the Fano effect,\cite{Fano2024,interf2024} a dip-peak feature can be observed at the resonance $\mu_{L}=\epsilon_4$. It is a manifestation of the quantum interference between the $\uparrow$ state with energy $\epsilon_4$ and $\downarrow$ state with energy $\epsilon_3\approx 0.279\,\epsilon_0$. An electron spin from the left lead can perform elastic tunneling through the $\uparrow$ state with energy $\epsilon_4$, or it can tunnel via a spin flip from the $\downarrow$ state with energy $\epsilon_3$ to the $\uparrow$ state with energy $\epsilon_4$, involving absorption of an energy $\omega\approx 0.371\,\epsilon_0$. 
Similarly, the tunneling can occur elastically via the $\downarrow$ state with energy $\epsilon_3$, or via a spin flip from the $\uparrow$ state with energy $\epsilon_4$ to the $\downarrow$ state with energy $\epsilon_3$, involving emission of an energy $\omega$. The two tunneling pathways, one elastic and the other inelastic involving a spin flip, ending up in the same final $\uparrow$ state with energy $\epsilon_{4}$, or the $\downarrow$ state with energy $\epsilon_{3}$, destructively (dip) or constructively (peak) interfere. The asymmetric line shape in $S^{zz}_{LL}$ (black line in Fig.~3) mimics the asymmetric Fano resonance profile.\cite{Fano2024,interf2024} For $\mu_{L}=\epsilon_4$, the corresponding resonant anisotropy parameter equals $D_{\rm res}\approx 0.00129\,\epsilon_0$ (see the grid line in Fig.~3). The destructive quantum interference corresponds to the dip (interference minimum) at $D_{\rm min}\approx 0.00101\,\epsilon_0$, while the constructive interference corresponds to the peak (interference maximum) at $D_{\rm max}\approx 0.00157\,\epsilon_0$. Around $D_{\rm res}$, the spin-current noise $S^{zz}_{LL}$ matches the Fano-like shape $\sigma^{z}_{S}(D)$ (red line in Fig.~3) expressed as
\begin{equation}
	\sigma^{z}_{S}(D)=C+A\frac{(D-D_{\rm res}+q\Gamma_{\rm res}/2)^2}{(D-D_{\rm res})^2+(\Gamma_{\rm res}/2)^2},\label{eq: Fano_D_res}
\end{equation}	
where $\Gamma_{\rm res}=|D_{\rm max}-D_{\rm min}|\approx 0.00056\,\epsilon_0$ is the width of the resonance. The asymmetry parameter $q=1$ shows that elastic and inelastic spin-flip processes occur with equal probability, so that $D_{\rm res}=(D_{\rm min}+D_{\rm max})/2$. Half of the amplitude of the Fano-like shape is given by $A=[S^{zz}_{LL}(D_{\rm max})-S^{zz}_{LL}(D_{\rm min})]/2\approx 0.000194\,\epsilon_0$, while $C=S^{zz}_{LL}(D_{\rm min})\approx 0.00229\,\epsilon_0$. The inset in Fig.~3 depicts individual contributions of two interference profiles around $D_{\rm res}$ to the Fano-like shape in $S^{zz}_{LL}$. One profile that includes interference of elastic and spin-flip pathways ending in the $\uparrow$ state with energy $\epsilon_4$ is presented in $S^{\uparrow\uparrow}_{LLS}-S^{\uparrow\downarrow}_{LLS}$ (pink line, inset in Fig.~3). The other profile with interfering elastic and spin-flip pathways ending in the $\downarrow$ state with energy $\epsilon_3$ is given by the contribution $S^{\downarrow\downarrow}_{LLS}-S^{\downarrow\uparrow}_{LLS}$ (purple line, inset in Fig.~3). The asymmetry parameter $q=1$ in both interference profiles, showing equal probabilities of elastic and spin-flip processes involving energy absorption (pink line) or emission (purple line).

The spin current $I_{Lz}$ and shot noise $S^{zz}_{LL}$ as functions of the chemical potential $\mu=\mu_{L}=\mu_{R}$ at zero temperature are presented in Figs.~4(a) and 4(b) for several values of the magnetic anisotropy parameter $D$. Since  $I^{\uparrow}_{Lz}$=$-I^{\downarrow}_{Lz}$ at zero-bias conditions, the corresponding charge current equals zero, with positive charge-current noise in the regions between levels connected with spin-flip events.\cite{4rad} In the same regions, the shot noise of spin current $S^{zz}_{LL}$ is positive, while the spin current $I_{Lz}$ takes positive (negative) values for $D<\omega_{L}/2S_z$ ($D>\omega_{L}/2S_z$). With the decrease of $|\omega|$, the magnitudes of $I_{Lz}$ and $S^{zz}_{LL}$ decrease. In Fig.~4(c) the dependence of spin-current shot noise $S^{zz}_{LL}$ on the bias voltage $eV=\mu_{L}-\mu_{R}$, with $\mu_{L,R}=\pm eV/2$, is plotted for different values of the anisotropy parameter $D$ at zero temperature. The positions of steps and dip-peak features in $S^{zz}_{LL}$ at values of $eV$ such that $\pm eV/2=\mu_{\xi}=\epsilon_i$ denote Floquet quasienergy levels $\epsilon_i$ available for spin transport. Thus they depend on the anosotropy parameter $D$. 
For large values of $|eV|$, the spin-current noise $S^{zz}_{LL}$ is saturated.

The Gilbert damping coefficient $\alpha$, the coefficient $\beta$, the magnitude of the in-plane torque component $T_{\bot}$, and the torque along the $z$ direction, $T_z$, are shown as functions of the magnetic anisotropy parameter $D$ in Fig.~5 for five different tilt angles $\theta$ at zero temperature. Starting with $\theta=0$, when $S=S_z$ (purple dotted lines in Fig.~5), one notices that $\alpha$ has two local maximums and two local minimums around values of $D$ that correspond to resonances $\mu_{\xi}=\epsilon_i$. The coefficient $\beta$ has four minimums at the same values of $D$. Taking into account that the spin of the molecule is static for $\theta=0$, with $\dot{\vec{S}}(t)=0$, and $\gamma =0$, according to Eqs.~(23) and (32), the torque components $T_{z}=0$ and $T_{\bot}=0$. Note the presence of the elastic spin currents here, $I_{Lz}=-I_{Rz}$ [see Eq.~(A3) in the Appendix]. 
On the other hand, for $\theta=\pi/2$, the molecular spin is perpendicular to the $z$-axis, $S_{z}=0$, and $\alpha$, $\beta$, $T_\bot$, and $T_z$ are independent of $D$ (pink dashed lines in Fig.~5). Grid lines in  Fig.~5 are related to $\theta=\pi/3$, $\omega_{L}=0.5\,\epsilon_0$ (green lines), and the values of parameter $D$, such that $\mu_\xi=\epsilon_i$. Around $D=-0.01312\,\epsilon_0$ corresponding to $\mu_{L}=\epsilon_2$ (grid line), one notices a local maximum in $\alpha$ and $T_{\bot}$, a local minimum in $\beta$, while $T_z$ has a minimum negative value. For this set of parameters, $\downarrow$ level $\epsilon_1$ and $\uparrow$ levels $\epsilon_2$ and $\epsilon_4$ lie within the bias-voltage window, while $\downarrow$ level $\epsilon_{3}<\mu_R$. This means that there are more inelastic tunneling pathways, involving electron spin-flip, available for $\downarrow$ electrons than for $\uparrow$ electrons. Namely, a tunneling electron in the $\downarrow$ level $\epsilon_3$ can flip its spin and enter the $\uparrow$ level $\epsilon_4$ within the bias voltage window, before it tunnels off the orbital, but the $\uparrow$ electron from the level $\epsilon_4$ cannot flip its spin and enter the $\downarrow$ level $\epsilon_3$, since $\epsilon_3$ lies below the bias voltage window. As a consequence, more $\downarrow$ electrons participate in inelastic spin transport, resulting in the negative torque component $T_z$. 
With further increase of $D$, for $\theta=\pi/3$ and $\omega_{L}=0.5\,\epsilon_0$, the torque $T_{z}<0$ until around $D=-0.00625\,\epsilon_0$, corresponding to $\mu_{R}=\epsilon_3$ (grid line), taking into account the level broadening $\Gamma$. Now all levels $\epsilon_i$  lie within the bias-voltage window, and $T_z$ has another local negative minimum and a steplike increase towards zero, while $\alpha$ has a local maximum and a steplike decrease towards zero. 
On the other hand, $T_{\bot}$ has a maximum value at $D=-0.00625$, while $\beta$ has another local minimum. As the increase of the anisotropy parameter $D$ continues, the Gilbert damping coefficient $\alpha$ and $T_z$ remain zero until around $D=0.00875\epsilon_0$, corresponding to $\mu_{R}=\epsilon_4$ (grid line). Here, $\vec{T}_z$ changes its direction with $T_z$ increasing towards a local maximum, while $\alpha$ reaches its maximum, $\beta$ has its minimum value, and $T_{\bot}$ has a local maximum. With further increase of $D$, levels $\epsilon_1$, $\epsilon_2$, and $\epsilon_3$ lie within the bias-voltage window, while the $\uparrow$ level $\epsilon_{4}<\mu_{R}$, and $\epsilon_{3}>\epsilon_{4}$ since $\omega<0$. Hence, there are more inelastic tunneling pathways available for $\uparrow$ particles, and the resulting $T_{z}>0$. Around $D=0.01406\,\epsilon_0$, corresponding to $\mu_{L}=\epsilon_1$ (grid line), the coefficient $\alpha$, the magnitude $T_{\bot}$, and $T_z$ have a local maximum, while $\beta$ has a local minimum for $\theta=\pi/3$ and $\omega_{L}=0.5\,\epsilon_0$ (green line in Fig.~5). Note that $T_{\bot}$ decreases from its maximum value to zero in the region $-0.00625\,\epsilon_{0}<D<0.005\,\epsilon_0$ ($\omega>0$), and increases for $0.005\,\epsilon_{0}<D<0.00875\,\epsilon_0$ ($\omega<0$). At $D=0.005\,\epsilon_0$, the total frequency $\omega=\omega_{L}-2DS_{z}=0$ for $\theta=\pi/3$ and $\omega_{L}=0.5\,\epsilon_0$, and the direction of the precession changes, while $T_{\bot}=0$. Similarly, for $\theta=\pi/9$ and $\theta=\pi/4$, the torque $T_{\bot}=0$ at $D=\omega_{L}/2S_z$, while $\alpha$ and $T_z$ vanish for the values of $D$ such that each $\epsilon_i$ satisfies $\mu_{R}\leq\epsilon_{i}\leq\mu_{L}$ with respect to level broadening $\Gamma$ (red and blue dot-dashed lines in Fig.~5). In the absence of the magnetic field ($\omega_{L}=0$), for $\theta=\pi/3$, the magnitude $T_{\bot}=0$ at $D=0$ [orange line in Fig.~5(c)]. At a high anisotropy, $\lvert D\rvert\gg\omega_{L}/2S_{z}$, the elastic tunneling processes are dominant and consequently $\alpha$, $\beta$, $T_{\bot}$, and $T_z$ approach zero.

The torque coefficients $\alpha$ and $\beta$, the magnitude $T_{\bot}$, and $T_z$ are presented as functions of the applied bias voltage $eV$ in Fig.~6 for several different magnetic anisotropy parameters $D$ at zero temperature. The bias voltage is varied according to $\mu_{L,R}=\pm eV/2$. For $D=\omega_L/2S_z=0.005\,\epsilon_0$, the molecular spin is static ($\omega=0$), and only elastic spin transport occurs through two available transport channels, so that $\alpha$, $\beta$, and $\vec{T}(t)$ vanish (purple dotted lines in Fig.~6). 
For $D\neq 0.005\,\epsilon_0$, one notices that $\alpha$ and $T_z$ approach constant values for chemical potentials $\mu_{\xi}$ positioned between quasienergy levels $\epsilon_i$ connected with spin-flip events. With the decrease of $D$, for $D<\omega_L/2S_z$, these values of $\alpha$ and $T_z$ decrease [pink dashed, red, and blue dot-dashed lines in Figs.~6(a) and 6(d)], vanish for $\omega=0$, and rise again for $D>\omega_L/2S_z$  [green lines in Figs.~6(a) and 6(d)]. Note that $T_{z}>0$ for $D<\omega_L/2S_z$, and $T_{z}<0$ for $D>\omega_L/2S_z$. Moreover, the width of the bias-voltage regions where the inelastic tunneling events occur depends on $D$ as $w=2\omega=2\omega_{L}-4DS_z$, taking into account the level broadening $\Gamma$. Within these regions, the coefficient $\beta$ increases(decreases) from a local minimum(maximum) to a local maximum(minimum), so that the torque $\beta\dot{\vec{S}}$ tends to oppose(enhance) the rotational motion of the molecular spin $\vec{S}$, for negative(positive) $\beta$. All peaks in $T_{\bot}$ correspond to $\mu_{\xi}=\epsilon_i$. Their values decrease with the increase of $D$ for $\omega>0$, drop to zero for $\omega=0$, and rise again for $\omega<0$ [see Fig.~6(c)]. For the values of $eV$ such that $\beta=0$, while $\alpha$ and $\lvert T_{z}\rvert$ approach their maximums, the magnitude $T_{\bot}$ has local minimums.
 
In Fig.~7 the coefficients $\alpha$ and $\beta$, the magnitude of the in-plane torque component, $T_{\bot}$, and $T_z$ are plotted as functions of the Larmor frequency $\omega_L$ at zero temperature for four different values of the magnetic anisotropy parameter $D$. The bias voltage is varied as $eV=\mu_L-\mu_R$, with $\mu_L=2.5\,\epsilon_0$ and $\mu_R=0$. The positions and values of all local maximums in $\alpha$, steps in $T_z$, peaks and dips in $\beta$, and peaks in $T_{\bot}$ depend on the parameter $D$. They correspond to resonances $\mu_{\xi}=\epsilon_i$. For the anisotropic molecular spin, $\alpha$, $\beta$, and $T_{\bot}$ are even functions, while $T_z$ is an odd function of $\omega$ ($\omega_L$ for $D=0$)\cite{we2013}.
Around $\omega_{L}=0.5\,\epsilon_0$ (grid lines), a maximum value for $D=0.00875\,\epsilon_0$ and a local maximum for $D=-0.00625\,\epsilon_0$ can be observed in $\alpha$ due to resonances $\mu_{R}=\epsilon_4$ and $\mu_{R}=\epsilon_3$, while $\beta$ shows negative minimums, $T_{\bot}$ local maximums, and $T_z$ steplike decreases (blue dot-dashed and green lines in Fig.~7). Looking at $\omega_{L}>0$ for $D=-0.00625\,\epsilon_0$ (green lines in Fig.~7), one notices that for $\omega_{L}<0.5\,\epsilon_0$ all quasienergy levels $\epsilon_i$ lie within the bias-voltage window. 
With the increase of $\omega_{L}$ they leave the bias-voltage window after reaching a resonance with $\mu_R$ or $\mu_L$: $\epsilon_{3}=\mu_{R}$ at $\omega_{L}=0.5\,\epsilon_0$, $\epsilon_{2}=\mu_{L}$ at $\omega_{L}=1.5\,\epsilon_0$, $\epsilon_{1}=\mu_{R}$ at $\omega_{L}=2.25\,\epsilon_0$, and $\epsilon_{4}=\mu_{L}$ at $\omega_{L}=3.25\,\epsilon_0$. If we take into account the level broadening $\Gamma$, all quasienergy levels $\epsilon_i$ leave the bias-voltage window around $\omega_{L}=3.25\,\epsilon_0$, with $\epsilon_{2(4)}>\mu_L$ and $\epsilon_{1(3)}<\mu_R$. 
With a further increase of $\omega_{L}$, $\alpha$ and $\beta$ vanish, while $T_{\bot}$ and $T_z$ become saturated due to spin-flip processes involving electron spins from the leads, entering $\downarrow$ levels $\epsilon_{1}$ or $\epsilon_3$, and absorbing energy $\omega$ during the interaction with the molecular spin (note that saturated $T_{z}<0$ for $\omega_{L}>0$).
At $\omega_{L}=2DS_{z}$, the STT vanishes. For example, at $\omega_{L}=0.5\,\epsilon_0$ for $D=0.005\,\epsilon_0$, $T_{\bot}=0$ and $T_{z}=0$ [intersection between grid line and pink-dashed line in Figs.~7(c) and 7(d)]. 

\section{Conclusions}

In this paper, the properties of spin transport through a junction consisting of a single orbital of a magnetic molecule with precessing anisotropic spin in a constant magnetic field were theoretically studied. The orbital is connected to two Fermi leads. The precession is kept undamped by external means, and the precession frequency involves two contributions: one is the Larmor frequency and the other is a term involving the uniaxial magnetic anisotropy parameter. Using the Keldysh NEGF technique, the spin currents, noise of $z$-polarized spin current, STT with the Gilbert damping coefficient, and other torque coefficients were derived.

The results were discussed and analyzed at zero temperature. The observed spin-transport characteristics such as steps, peaks, and dips are related to resonances between chemical potentials of the leads and molecular quasienergy levels, which depend on the uniaxial magnetic anisotropy parameter of the molecular spin. 
The inelastic tunneling of an electron spin that involves a spin-flip and absorption (emission) of an energy that depends on the magnetic anisotropy occurs due to exchange interaction with the precessing anisotropic molecular spin. 
During that process, the STT is exerted on the spin of the molecule and compensated by external means. The opposite torque is exerted by the molecular spin onto the spin currents, thus affecting the spin-transport properties of the junction.
Both torque coefficients $\alpha$ and $\beta$, as well as the magnitude of the in-plane torque $T_{\bot}$, are even functions, while $T_z$ is an odd function of the total precession frequency, taking into consideration the contribution of the magnetic anisotropy to the precession.

By adjusting the uniaxial magnetic anisotropy parameter with respect to the Larmor precession frequency and the tilt angle between the magnetic field and the molecular spin, one can control the spin current and noise, the STT, the Gilbert damping, and the other torque coefficients.
The quantum interference effects between the states connected with spin-flip processes are manifested in the spin-current noise as peaks and dips, resembling the Fano-like resonance profiles, controlled by the anisotropy parameter and Larmor frequency. 
Furthermore, it might be possible to perform a measurement of the dc-spin current and STT components, since they reveal the quasienergy level structure in the molecular orbital. All spatial components of the STT vanish for the anisotropy contribution to the precession frequency that matches the Larmor frequency, when the precession is suppressed, thus allowing to determine the anisotropy and other parameters. Also, the large magnetic anisotropy parameter leads to vanishing of the spin current, spin-current noise, and STT. 

Taking into account that the spin currents and noise, and the magnitude, direction, and sign of the STT, can be manipulated by the uniaxial magnetic anisotropy and other parameters, even if the magnetic field is turned off, the obtained results might be useful in spintronic applications using molecular magnets. Since the magnetic anisotropy parameter is important for applications of molecular magnets in magnetic storage, it might be suitable to study the noise of STT to obtain additional characteristics of spin-transport, and to use a quantum-mechanical description of the anisotropic molecular spin in tunnel junctions with normal or ferromagnetic leads.

\begin{acknowledgments}
	The author acknowledges funding provided by the Institute of Physics Belgrade, through the Grant No. 451-03-68/2022-14/200024 of the Ministry of Education, Science, and Technological Development of the Republic of Serbia.
\end{acknowledgments}

\begin{widetext}
	
	\appendix*
	
	\section*{Appendix: Expressions for the spin-current spatial components and zero-frequency limit of the torque coefficients}
	\renewcommand{\theequation}{A\arabic{equation}}
	\setcounter{equation}{0}
	
The expressions for complex functions $I_{\xi x}(D)$ and $I_{\xi y}(D)$ introduced by Eqs.~(7) and (8), and spin-current component $I_{\xi z}$, are presented here in terms of the Green's functions $\hat{G}^{0r}(\epsilon)$ and $\hat{G}^{0a}(\epsilon)=[\hat{G}^{0r}(\epsilon)]^\dagger$. They can be written as
	\begin{align}
		I_{\xi x}(D)=\label{eq: x resonant}&-i\int\frac{d\epsilon}{4\pi}\Bigg\{\frac{\Gamma_{\xi}\Gamma_{\eta}}{\Gamma}[f_{\xi}(\epsilon)-f_{\eta}(\epsilon)]
		\Bigg[\frac{\gamma G^{0r}_{11}(\epsilon+\omega_{L}-2DS_{z})G^{0r}_{22}(\epsilon)}
		{\lvert 1-\gamma^{2}G^{0r}_{11}(\epsilon+\omega_{L}-2DS_{z})G^{0r}_{22}(\epsilon)\lvert^{2}}\nonumber\\
		&+\frac{2i\gamma{\rm Im}
			\{G^{0r}_{11}(\epsilon)\}G^{0a}_{22}(\epsilon-\omega_{L}+2DS_{z})+\gamma^{3}\lvert G^{0r}_{11}(\epsilon) G^{0r}_{22}(\epsilon-\omega_{L}+2DS_{z})\lvert^{2}}
		{\lvert 1-\gamma^{2}G^{0r}_{11}
			(\epsilon)G^{0r}_{22}(\epsilon-\omega_{L}+2DS_{z})\lvert^{2}}\Bigg]\nonumber\\
		&+\sum_{\lambda,\zeta=L,R}\frac{\Gamma_{\lambda}\Gamma_{\zeta}}{\Gamma}[f_{\lambda}(\epsilon-\omega_{L}+2DS_{z})-f_{\zeta}(\epsilon)]\nonumber\\
		&\times\Big [\delta_{\zeta\xi}-\delta_{\lambda\xi}\gamma^{2}G^{0a}_{11}(\epsilon)G^{0r}_{22}(\epsilon-\omega_{L}+2DS_{z})\Big ]
		\frac{\gamma G^{0r}_{11}(\epsilon)G^{0a}_{22}(\epsilon-\omega_{L}+2DS_{z})}
		{\lvert 1-\gamma^{2}G^{0r}_{11}(\epsilon)G^{0r}_{22}(\epsilon-\omega_{L}+2DS_{z})\lvert^{2}}\Bigg \},\,\,\eta\neq\xi,\\
		I_{\xi y}(D)=\label{eq: y resonant}&\,\ iI_{\xi x}(D),\\ 
		{\rm and}\,\ I_{\xi z}=\label{eq: z resonant}&\int\frac{d\epsilon}{4\pi}\Bigg\{\frac{\Gamma_{\xi}\Gamma_{\eta}}{\Gamma}[f_{\xi}(\epsilon)-f_{\eta}(\epsilon)]
		\Bigg[\frac{2{\rm Im}\{G^{0r}_{11}(\epsilon)\}}
		{\lvert 1-\gamma^{2}G^{0r}_{11}(\epsilon)G^{0r}_{22}(\epsilon-\omega_{L}+2DS_{z})\lvert^{2}}-\frac{2{\rm Im}\{G^{0r}_{22}(\epsilon)\}}
		{\lvert 1-\gamma^{2}G^{0r}_{11}(\epsilon+\omega_{L}-2DS_{z})G^{0r}_{22}(\epsilon)\lvert^{2}}\Bigg]\nonumber\\
		&+\sum_{\lambda,\zeta=L,R}\Gamma_{\lambda}\Gamma_{\zeta}[f_{\lambda}(\epsilon-\omega_{L}+2DS_{z})-f_{\zeta}(\epsilon)]
		(\delta_{\lambda\xi}+\delta_{\zeta\xi})
		\frac{\gamma^{2}\lvert G^{0r}_{11}(\epsilon) G^{0r}_{22}(\epsilon-\omega_{L}+2DS_{z})\lvert^{2}}
		{\lvert 1-\gamma^{2}G^{0r}_{11}(\epsilon)G^{0r}_{22}(\epsilon-\omega_{L}+2DS_{z})\lvert^{2}}\Bigg \},\,\eta\neq\xi.
	\end{align}

In the limit $\omega\rightarrow 0$, i.e., $D\rightarrow\omega_{L}/2S_z$, considering that $f_{\xi}(\epsilon-\omega)-f_{\xi}(\epsilon)\rightarrow -\omega\partial_{\epsilon}f_{\xi}(\epsilon)$, the expressions for the spin-torque coefficients $\alpha$ and $\beta$ given by Eqs.~(29) and (30) result in 

	\begin{align}
	\lim_{\omega\to 0}\alpha=\label{eq: alpha_limit}&\frac{\Gamma}{S}\int\frac{d\epsilon}{2\pi}\sum_{\xi}\Gamma_{\xi}
	[\partial_{\epsilon}f_{\xi}(\epsilon)]\frac{( JS_{z}/2\Gamma){\rm Im}\{G^{0r}_{11}(\epsilon)G^{0a}_{22}(\epsilon)\}-\gamma^{2}\lvert G^{0r}_{11}
		(\epsilon)G^{0r}_{22}(\epsilon)\lvert^2}
	{\lvert 1-\gamma^{2}G^{0r}_{11}(\epsilon)G^{0r}_{22}(\epsilon)\lvert^{2}},\\
	\lim_{\omega\to 0}\beta=\label{eq: beta_limit}&J\int\frac{d\epsilon}{4\pi}\sum_{\xi}\Gamma_{\xi}
	[\partial_{\epsilon}f_{\xi}(\epsilon)]\frac{{\rm Re}\{G^{0r}_{11}(\epsilon)G^{0a}_{22}(\epsilon)\}-\gamma^{2}\lvert G^{0r}_{11}(\epsilon) G^{0r}_{22}(\epsilon)\lvert^2}
	{\lvert 1-\gamma^{2}G^{0r}_{11}(\epsilon)G^{0r}_{22}(\epsilon)\lvert^{2}}.
\end{align}

\end{widetext}


\begin{thebibliography}{12}

\bibitem{m1}
L. Thomas, F. Lionti, R. Ballou, D. Gatteschi, R. Sessoli, and B. Barbara, Macroscopic quantum tunnelling of magnetization in a single crystal of nanomagnets, Nature (London) {\bf 383}, 145-147 (1996).

\bibitem{m2} 
D. Gatteschi, R. Sessoli, and J. Villain, \textit{Molecular Nanomagnets}, Oxford University Press, New York (2006).

\bibitem{m3} 
L. Bogani and W. Wernsdorfer, Molecular spintronics using single-molecule magnets, Nature Mater. {\bf 7}, 179-186 (2008).

\bibitem{m4}
C. Timm and M. Di Ventra, Memristive properties of single-molecule magnets, Phys. Rev B {\bf 86}, 104427 (2012).

\bibitem{m5}
M. N. Leuenberger and D. Loss, Quantum computing in molecular magnets, Nature (London) {\bf 410}, 789-793 (2001).

\bibitem{m6}
R. E. P. Winpenny, Quantum information processing using molecular nanomagnets as qubits, Angew. Chem. Int. Ed., {\bf 47}, 7992-7994 (2008).

\bibitem{new1}
M. -H. Jo, J. E. Grose, K Baheti, M. M. Deshumukh, J. J. Sokol, E. M. Rumberger, D. N. Hendrickson, J. R. Long, H. Park, and D. C. Ralph, Signatures of molecular magnetism in single-molecule transport spectroscopy, Nano Lett. {\bf 6}, 2014 (2006).

\bibitem{coulombblockade2}
A. S. Zyazin, J. W. G. van den Berg, E. A. Osorio, H. S. J. van der Zant, N. P. Konstantinidis, M. Leijnse, M. R. Wegewijs, F. May, W. Hofstetter, C. Danieli, and A Cornia, Electric field controlled magnetic anisotropy in a single molecule, Nano Lett. {\bf 10}, 3307 (2010).

\bibitem{ele1}
E. Burzur\'{i}, A. S. Zyazin, A. Cornia, and H. S. J. van der Zant, Direct Observation of Magnetic Anisotropy in an Individual $\rm Fe_4$ Single-Molecule Magnet, Phys. Rev. Lett. {\bf 109}, 147203 (2012).

\bibitem{t1}
M. Misiorny, M. Hell, and M. R. Wegewijs, Spintronic magnetic anisotropy, Nature Physics {\bf 9}, 801-805 (2013).

\bibitem{new5}
H. B. Heersche, Z. de Groot, J. A. Folk, H. S. J. van der Zant, C. Romeike, M.R. Wegewijs, L. Zobbi, D. Barreca, E. Tondello, and A. Cornia, Electron transport through single $\rm Mn_{12}$ molecular magnets, Phys. Rev. Lett. {\bf 96}, 206801 (2006).

\bibitem{new6}
R. Sessoli, Magnetic molecules back in the race, Nature {\bf 548}, 400-401 (2017).

\bibitem{new7}
F. Delgado and J. Fern\'andez-Rossier, Storage of Classical Information in Quantum Spins, Phys. Rev. Lett. {\bf 108}, 196602 (2012).

\bibitem{new8}
S. Kahle, Z. Deng, N. Malinowski, C. Tonnoir, A. Forment-Aliaga, N. Thontasen, G. Rinke, D. Le, V. Turkowski, T. S. Rahman, S. Rauschenbach, M. Ternes, and K. Kern, The Quantum Magnetism of Individual Manganese-12-Acetate Molecular Magnets Anchored at Surfaces, Nano Lett. {\bf 12}, 518-521 (2012). 

\bibitem{new9}
F. -S. Guo, B. M. Day, Y. -C. Chen, M. -L. Tong, A. Mansikkam\"{a}ki, and R. A. Layfield, Magnetic hysteresis up to 80 kelvin in a dysprosium metallocene single-molecule magnet, Science {\bf 362}, 1400-1403 (2018).

\bibitem{m7}
R. Sessoli, D. Gatteschi, A. Caneschi, and M. A. Novak, Magnetic bistability in a metal-ion cluster, Nature (London) {\bf 365} 141-143 (1993).

\bibitem{m8}
M. Misiorny and J. Barna\'{s}, Effects of intrinsic spin-relaxation in molecular magnets on current-induced magnetic switching, Phys. Rev. B {\bf 77}, 172414 (2008).

\bibitem{an1}
Y. Shiota, T. Nozaki, F. Bonell, S. Murakami, T. Shinjo, and Y. Suzuki, Induction of coherent magnetization switching in a few atomic layers of FeCo using voltage pulses, Nature Mater. {\bf 11}, 39-43 (2012).

\bibitem{an2}
B. W. Heinrich, L. Braun, J. I. Pascual, K. J. Franke, Tuning the magnetic anisotropy of single molecules, Nano Lett. {\bf 15}, 4024-4028 (2015).

\bibitem{an3}
J. D. V. Jaramillo, H. Hammar, and J. Fransson, Electronically mediated magnetic anisotropy in vibrating magnetic molecules, ACS Omega {\bf 3}, 6546-6553 (2018).

\bibitem{an4}
B. Rana and Y. Otani, Towards magnonic devices based on voltage-controlled magnetic anisotropy, Commun. Phys. {\bf 2}, 90 (2019).

\bibitem{ele2}
R. E. George, J. P. Edwards, and A. Ardavan, Coherent spin control by electrical manipulation of the magnetic anisotropy, Phys. Rev. Lett. {\bf 110}, 027601 (2013).

\bibitem{ele3}
A. Sarkar and G. Rajaraman, Modulating magnetic anisotropy in Ln(III) single-ion magnets using an external electric field, Chem. Sci. {\bf 11}, 10324-10330 (2020).

\bibitem{ele4}
Y. Lu, Y. Wang, L. Zhu, L. Yang, and L. Wang, Electric field tuning of magnetic states in single magnetic molecules, Phys. Rev. B {\bf 106}, 064405 (2022).

\bibitem{new10}
M. Pletyukhov, D. Schuricht, and H. Schoeller, Relaxation versus Decoherence: Spin and Current Dynamics in the Anisotropic Kondo Model at Finite Bias and Magnetic Field, Phys. Rev. Lett. {\bf 104}, 106801 (2010).

\bibitem{new11}
R. Wieser, Comparison of Quantum and Classical Relaxation in Spin Dynamics, Phys. Rev. Lett. {\bf 110}, 147201 (2013).

\bibitem{new12}
R. Mondal, M. Berritta, and P. M. Oppeneer, Relativistic theory of spin relaxation mechanisms in the Landau-Lifshitz-Gilbert equation of spin dynamics, Phys. Rev. B {\bf 94}, 144419 (2016).

\bibitem{new13}
M. Sayad, R. Rausch, and M. Potthoff, Relaxation of a Classical Spin Coupled to a Strongly Correlated Electron System, Phys. Rev. Lett. {\bf 117}, 127201 (2016).

\bibitem{new14}
A. Barman, S. Mandal, S. Sahoo, and A. De, Magnetization dynamics of nanoscale magnetic materials: A perspective, J. Appl. Phys. {\bf 128}, 170901 (2020).

\bibitem{new15}
L. Yang, P. Glasenapp, A. Greilich, D. Reuter, A. D. Wieck, D. R. Yakovlev, M. Bayer, and S. A. Crooker, Two-colour spin noise spectroscopy and fluctuation correlations reveal homogeneous linewidths within quantum-dot ensembles, Nat. Commun. {\bf 5}, 4949 (2014).

\bibitem{new16}
W. D. Rice, W. Liu, T. A. Baker, N. A. Sinitsyn, V. I. Klimov, and S. A. Crooker, Revealing giant internal magnetic fields due to spin fluctuations in magnetically doped colloidal nanocrystals, Nat. Nanotechnol. {\bf 11}, 137-142 (2016).

\bibitem{new17}
M. Swar, D. Roy, S. Bhar, S. Roy, and S. Chaudhuri, Detection of spin coherence in cold atoms via Faraday rotation fluctuations, Phys. Rev. Res. {\bf 3}, 043171 (2021).

\bibitem{new18}
C. Stahl and M. Potthoff, Anomalous Spin Precession under a Geometrical Torque, Phys. Rev. Lett {\bf 119}, 227203 (2017).

\bibitem{new19}
U. Bajpai and B. K. Nikoli\'{c}, Spintronics Meets Nonadiabatic Molecular Dynamics: Geometric Spin Torque and Damping on Dynamical Classical Magnetic Texture due to an Electronic Open Quantum System, Phys. Rev. Lett. {\bf 125}, 187202 (2020).

\bibitem{new191}
P. M. Gunnink, T. Ludwig, and R. A. Duine, Charge conservation in spin-torque oscillators leads to a self-induced torque, Phys. Rev. B {\bf 109}, 024408 (2024).

\bibitem{new20}
D. C. Ralph and M. D. Stiles, Spin transfer torques, J. Magn. Magn. Mater. {\bf 320}, 1190-1216 (2008).

\bibitem{new211}
J.-X. Zhu, Z. Nussinov, A. Shnirman, and A. V. Balatsky, Novel Spin Dynamics in a Josephson Junction, Phys. Rev. Lett. {\bf 92}, 107001 (2004).

\bibitem{new212}
Z. Nussinov, A. Shnirman, D. P. Arovas, A. V. Blaltsky, and J.-X. Zhu, Spin and spin-wave dynamics in Josephson junctions, Phys. Rev. B {\bf 71}, 214520 (2005).

\bibitem{new213}
J.-X. Zhu and J. Fransson, Electric field control of spin dynamics in a magnetically active tunnel junction, J. Phys.: Condens. Matter {\bf 18}, 9929-9936 (2006).

\bibitem{new214}
J. Fransson and J.-X. Zhu, Spin dynamics in a tunnel junction between ferromagnets, New J. Phys. {\bf 10}, 013017 (2008).

\bibitem{new21}
J. Fransson, Subnanosecond switching of local spin-exchange coupled to ferromagnets, Phys. Rev. B {\bf 77}, 205316 (2008).

\bibitem{new22}
J. Fransson. J. Ren, and J.-X. Zhu, Electrical and thermal control of magnetic exchange interactions, Phys. Rev. Lett. {\bf 113}, 257201 (2014).

\bibitem{new23}
H. Hammar and J. Fransson, Time-dependent spin and transport properties of a single-molecule magnet in a tunnel junction, Phys. Rev. B {\bf 94}, 054311 (2016).

\bibitem{new24}
H. Hammar and J. Fransson, Transient spin dynamics in a single-molecule magnet, Phys. Rev. B {\bf 96}, 214401 (2017).

\bibitem{new25}
H. Hammar and J. Fransson, Dynamical exchange and phase induced switching of a localized molecular spin, Phys. Rev. B {\bf 98}, 174438 (2018).

\bibitem{new26}
H. Hammar, J. D. V. Jaramillo, and J. Fransson, Spin-dependent heat signatures of single-molecule spin dynamics, Phys. Rev. B {\bf 99}, 115416 (2019).

\bibitem{new27}
I. Makhfudz, E. Olive, and S. Nicolis, Nutation wave as a platform for ultrafast spin dynamics in ferromagnets, Appl. Phys. Lett. {\bf 117}, 132403 (2020).

\bibitem{new28}
R. Rahman and S. Bandyopadhyay, An observable effect of spin inertia in slow magneto-dynamics: increase of the switching error rates in nanoscale ferromagnets, J. Phys.: Condens. Matter {\bf 33}, 355801 (2021).

\bibitem{new29}
J. C. Slonczewski, Current-driven excitation of magnetic multilayers, J. Magn. Magn. Mater. {\bf 159}, L1-L7 (1996).

\bibitem{new30}
L. Berger, Emission of spin waves by a magnetic multilayer traversed by a current, Phys. Rev. B {\bf 54}, 9353-9358 (1996).

\bibitem{new31}
M. Tsoi, A. G. M. Jansen, J. Bass, W. -C. Chiang, M. Seck, V. Tsoi and P. Wyder, Excitation of a Magnetic Multilayer by an Electric Current, Phys. Rev. Lett. {\bf 80}, 4281 (1998).

\bibitem{new32}
J. Z. Sun, Current-driven magnetic switching in manganite trilayer junctions, J. Magn. Magn. Mater. {\bf 202}, 157-162 (1999).

\bibitem{new33}
Y. Tserkovnyak, A. Brataas, and G. E. W. Bauer, Spin pumping and magnetization dynamics in metallic multilayers, Phys. Rev. B {\bf 66}, 224403 (2002).

\bibitem{new34}
G. E. W. Bauer, E. Saitoh, and B. J. van Wees, Spin caloritronics, Nature Mater. {\bf 11}, 391 (2012).

\bibitem{new35}
E. B. Myers, D. C. Ralph, J. A. Katine, R. N. Louie, and R. A. Buhrman, Current-Induced Switching of Domains in Magnetic Multilayer Devices, Science {\bf 285}, 867-870 (1999).

\bibitem{new36}
F. J. Albert, N. C. Emley, E. B. Myers, D. C. Ralph, and R. A. Buhrman, Quantitative Study of Magnetization Reversal by Spin-Polarized Current in Magnetic Multilayer Nanopillars, Phys. Rev. Lett. {\bf 89}, 226802 (2002).

\bibitem{new37}
S. I. Kiselev, J. C.Sankey, I. N. Krivorotov, N. C. Emley, R. J. Schoelkopf, R. A. Buhrman, and D. C. Ralph, Microwave oscillations of a nanomagnet driven by a spin-polarized current, Nature {\bf 425}, 380-383 (2003).

\bibitem{new38}
L. Liu, C. -F. Pai, Y. Li, H. W. Tseng, D. C. Ralph, and R. A. Buhrman, Spin-Torque Switching with the Giant Spin Hall Effect of Tantalum, Science {\bf 336}, 555-558 (2012).

\bibitem{new381}
A. Chudnovskiy, Ch. H\"{u}bner, B. Baxevanis, and D. Pfannkuche, Spin switching: From quantum to quasiclassical approach, Phys. Status Solidi B, {\bf 251}, 1764 (2014).

\bibitem{new382}
T. Ludwig, I. S. Burmistrov, Y. Gefen, and A. Shnirman, Strong nonequilibrium effects in spin-torque systems, Phys. Rev. B {\bf 95}, 075425 (2017).

\bibitem{new39}
M. Misiorny and J. Barna\'{s}, Magnetic switching of a single molecular magnet due to spin-polarized current, Phys. Rev. B {\bf 75}, 134425 (2007).

\bibitem{new40}
M. Misiorny and J. Barna\'{s}, Effects of Transverse Magnetic Anisotropy on Current-Induced Spin Switching, Phys. Rev. Lett. {\bf 111}, 046603 (2013).

\bibitem{new41}
K. Wrze\'{s}niewski and I. Weymann, Time-dependent spintronic anisotropy in magnetic molecules, Phys. Rev B {\bf 101}, 245434 (2020).

\bibitem{new411}
H.-B. Xue, J.-Q. Liang and W.-M. Liu, Manipulation and readout of spin states of a single-molecule magnet by a spin-polarized current, Physica E {\bf 138}, 115086 (2022).

\bibitem{new42}
C. Timm and F. Elste, Spin amplification, reading, and writing in transport through anisotropic magnetic molecules, Phys. Rev. B {\bf 73}, 235304 (2006).

\bibitem{new43}
K. Xia, P. J. Kelly, G. E. W. Bauer, A. Brataas, and I. Turek, Spin torques in ferromagnetic/normal-metal structures, Phys. Rev B {\bf 65}, 220401(R) (2002).

\bibitem{new44}
Y. Tserkovnyak, A. Brataas, G. E. E. Bauer, and B. I. Halperin, Nonlocal magnetization dynamics in ferromagnetic heterostructures, Rev. Mod. Phys. {\bf 77}, 1375 (2005).

\bibitem{new45}
E. M. Hankiewicz, G. Vignale, and Y. Tserkovnyak, Gilbert damping and spin Coulomb drag in a magnetized electron liquid with spin-orbit interaction, Phys. Rev. B {\bf 75}, 174434 (2007).

\bibitem{new46}
E. M. Hankiewicz, G. Vignale, and Y. Tserkovnyak, Inhomogeneous Gilbert damping from impurities and electron-electron interactions, Phys. Rev. B {\bf 78}, 020404(R) (2008).

\bibitem{new47}
Y. Tserkovnyak, E. M. Hankiewicz, and G. Vignale, Transverse spin diffusion in ferromagnets, Phys. Rev. B {\bf 79}, 094415 (2009).

\bibitem{new48}
G. Tatara, H. Kohno, and J. Shibata, Microscopic approach to current-driven domain wall dynamics, Phys. Rep. {\bf 468}, 213 (2008).

\bibitem{new49}
G. Tatara, Effective gauge field theory of spintronics, Physica E {\bf 106}, 208 (2019).

\bibitem{new50}
C. Bell, S. Milikisyants, M. Huber, and J. Aarts, Spin Dynamics in a Superconductor-Ferromagnet Proximity System, Phys. Rev. Lett. {\bf 100}, 047002 (2008).

\bibitem{new51}
M. Houzet, Ferromagnetic Josephson Junction with Precessing Magnetization, Phys. Rev. Lett. {\bf 101}, 057009 (2008).

\bibitem{new52}
J. P. Morten, A. Brataas, G. E. W. Bauer, W. Belzig, and Y. Tserkovnyak, Proximity-effect-assisted decay of spin currents in superconductors, EPL {\bf 84}, 57008 (2008).

\bibitem{new53}
S. Teber, C. Holmqvist, and M. Fogelst\"{o}rm, Transport and magnetization dynamics in a superconductor/single-molecule magnet/superconductor junction, Phys. Rev. B {\bf 81}, 174503 (2010).

\bibitem{new54}
 C. Holmqvist, M. Fogelst\"{o}rm, and W. Belzig, Spin-polarized Shapiro steps and spin-precession-assisted multiple Andreev reflection, Phys. Rev. B {\bf 90}, 014516 (2014).

\bibitem{new55}
Y. Yao, Q. Song, Y. Takarnura, J. P. Cascales, W. Yuan, Y. Ma, Y. Yun, X. C. Xie, J. S. Moodera, and W. Han, Probe of spin dynamics in superconducting NbN thin films via spin pumping, Phys. Rev. B {\bf 97}, 224414 (2018).

\bibitem{new56}
T. Kato, Y. Ohnuma, M. Matsuo, J. Rech, T. Jonckheere, and T. Martin, Microscopic theory of spin transport at the interface between a superconductor and a ferromagnetic insulator, Phys. Rev. B {\bf 99}, 144411 (2019).

\bibitem{Jauho1993}
N. S. Wingreen, A.-P. Jauho, and Y. Meir, Time-dependent transport through a mesoscopic structure, Phys. Rev. B {\bf 48}, 8487 (1993).

\bibitem{Jauho1994}
A.-P. Jauho, N. S. Wingreen, and Y. Meir, Time-dependent transport in interacting and noninteracting resonant-tunneling systems, Phys. Rev. B {\bf 50}, 5528 (1994).

\bibitem{JauhoBook}
A.-P. Jauho and H. Haug, \textit{Quantum Kinetics in Transport and Optics of
	Semiconductors} (Springer, Berlin, 2008).

\bibitem{new57}
Q.-f. Sun, H. Guo, and J. Wang, A Spin Cell for Spin Current, Phys. Rev. Lett. {\bf 90}, 258301 (2003).

\bibitem{new58}
J. Fransson and M. Galperin, Inelastic scattering and heating in a molecular spin pump, Phys. Rev. B {\bf 81}, 075311 (2010).

\bibitem{new59}
A. Saraiva-Souza, M. Smeu, L. Zhang, A. G. Souza Filho, H. Guo, and M. A. Ratner, Molecular Spintronics: Destructive Quantum Interference Controlled by a Gate, J. Am. Chem. Soc. {\bf 136}, 42, 15065-15071 (2014).

\bibitem{new60}
J. Fransson, O. Eriksson, and A. V. Balatsky, Theory of spin-polarized scanning tunneling microscopy applied to local spins, Phys. Rev. B {\bf 81}, 115454 (2010).

\bibitem{new61}
D. Rai and M. Galperin, Spin inelastic currents in molecular ring junctions, Phys. Rev. B {\bf 86}, 045420 (2012).

\bibitem{we2013}
M. Filipovi\'{c}, C. Holmqvist, F. Haupt, and W. Belzig, Spin transport and tunable Gilbert damping in a single-molecule magnet junction, Phys. Rev. B {\bf 87}, 045426 (2013); {\bf 88}, 119901 (2013).

\bibitem{we2016}
M. Filipovi\'{c} and W. Belzig, Photon-assisted electronic and spin transport in a junction containing precessing molecular spin, Phys. Rev. B {\bf 93}, 075402 (2016).

\bibitem{new62}
S. Lounis, A. Bringer, and S. Bl\"{u}gel, Magnetic Adatom Induced Skyrmion-Like Spin Texture in Surface Electron Waves, Phys. Rev. Lett {\bf 108}, 207202 (2012).

\bibitem{new63}
D. Yudin, D. R. Gulevich, and M. Titov, Light-Induced Anisotropic Skyrmion and Stripe Phases in a Rashba Ferromagnet, Phys. Rev. Lett. {\bf 119},147202 (2017).

\bibitem{new631}
F. Xu, G. Li, J. Chen, Z. Yu, L. Zhang, B. Wang, and J. Wang, Unified framework of the microscopic Landau-Lifshitz-Gilbert equation and its application to skyrmion dynamics, Phys. Rev. B {\bf 108}, 144409 (2023).

\bibitem{new64}
B. Wang, J. Wang, and H. Guo, Shot noise of spin current, Phys. Rev. B {\bf 69}, 153301 (2004).

\bibitem{new65}
F. M. Souza, A.-P. Jauho, and J. C. Egues, Spin-polarized current and shot noise in the presence of spin flip in a quantum dot via nonequilibrium Green's functions, Phys. Rev. B {\bf 78}, 155303 (2008).

\bibitem{we2018}
M. Filipovi\'{c} and W. Belzig, Shot noise of charge and spin transport in a junction with a precessing molecular spin, Phys. Rev. B {\bf 97}, 115441 (2018).

\bibitem{kondo1}
C. Romeike, M. R. Wegewijs, W. Hofstetter, and H. Schoeller, Quantum-tunneling-induced Kondo effect in single molecular magnets, Phys. Rev. Lett {\bf 96}, 196601 (2006).

\bibitem{stretching}
J. J. Parks, A. R. Champagne, T. A. Costi, W. W. Shum, A. N. Pasupathy, E. Neuscamman, S. Flores-Torres, P. S. Cornaglia, A. A. Aligia, C. A. Balseiro, G. K.-L. Chan, H. D. Abru\~{n}a, and D. C. Ralph, Mechanical control of spin states in spin-1 molecules and the underscreened Kondo effect, Science {\bf 328}, 1370-1373 (2010).

\bibitem{kondo2}
F. Elste and C. Timm, Resonant and Kondo tunneling through molecular magnets, Phys Rev. B {\bf 81}, 024421 (2010).

\bibitem{kondo3}
M. Misiorny, I. Weymann, and J. Barna\'s, Temperature dependence of electronic transport through molecular magnets in the Kondo regime, Phys. Rev. B {\bf 86}, 035417 (2012).

\bibitem{kondo4}
Y. Li, H. Kan, Y. Miao, S. Qiu, G. Zhang, J. Ren, C. Wang, and G. Hu, Magnetic manipulation of orbital hybridization and magnetoresistance in organic ferromagnetic co-oligomers, Physica E {\bf 124}, 114327 (2020).

\bibitem{seebeck1}
R.-Q. Wang, L. Sheng, R. Shen, B. Wang, and D. Y. Xing, Thermoelectric effect in single-molecule-magnet junctions, Phys. Rev. Lett. {\bf 105}, 057202 (2010).

\bibitem{seebeck2}
M. Misiorny and J. Barna\'{s}, Spin-dependent thermoelectric effects in transport through a nanoscopic junction involving a spin impurity, Phys. Rev. B {\bf 89}, 235438 (2014).

\bibitem{seebeck3}
M. Misiorny and J. Barna\'{s}, Effect of magnetic anisotropy on spin-dependent thermoelectric effects in nanoscopic systems, Phys. Rev. B {\bf 91}, 155426 (2015).

\bibitem{new69}
M. Misiorny and J. Barna\'{s}, Spin polarized transport through a single-molecule magnet: Current-induced magnetic switching, Phys. Rev. B {\bf 76}, 054448 (2007).

\bibitem{new70}
Z. Zhang and L. Jiang, Bias voltage induced resistance switching effect in single-molecule magnets' tunneling junction, Nanotechnology {\bf 25}, 365201 (2014).

\bibitem{spinblockade2}
A. P\l{}omi\'nska and I. Weymann, Pauli spin blockade in double molecular magnets, Phys. Rev. B {\bf 94}, 035422 (2016).

\bibitem{spinblockade3}
A. P\l{}omi\'nska and I. Weymann, Magnetoresistive properties of a double magnetic molecule spin valve in different geometrical arrangements, J. Magn. Magn. Mater. {\bf 480}, 11-21 (2019).

\bibitem{new71}
J. R. Petta, S. K. Slater, and D. C. Ralph, Spin-Dependent Transport in Molecular Tunnel Junctions, Phys. Rev. Lett. {\bf 93}, 136601 (2004).

\bibitem{new72}
U. Ham and W. Ho, Spin Splitting Unconstrained by Electron Pairing: The Spin-Vibronic States, Phys. Rev. Lett. {\bf 108}, 106803 (2012).

\bibitem{new73}
G. Czap, P. J. Wagner, F. Xue, L. Gu, J. Li, J. Yao, R. Wu, and W. Ho, Probing and imaging spin interactions with a magnetic single-molecule sensor, Science {\bf 364}, 670 (2019).

\bibitem{new74}
A. Landig, J. Koski, P. Scarlino, C. Reichl, W. Wegscheider, A. Wallraff, K. Ensslin, and T. Ihn, Microwave-Cavity-Detected Spin Blockade in a Few-Electron Double Quantum Dot, Phys. Rev. Lett. {\bf 122}, 213601 (2019).

\bibitem{new75}
M. Urdampilleta, S. Klyatskaya, J. -P. Cleuziou, M. Ruben, and W. Wernsdorfer, Supramolecular spin valves, Nature Mater. {\bf 10}, 502 (2011).

\bibitem{new76}
U. Ham and W. Ho, Imaging single electron spin in a molecule trapped within a nanocavity of tunable dimension, J. Chem. Phys. {\bf 138}, 074703 (2013).

\bibitem{new761}
L. Landau and E. M. Lifshitz, On the Theory of the Dispersion of Magnetic Permeability in Ferromagnetic Bodies, Phys. Z. Sowjetunion {\bf 8}, 153 (1935).

\bibitem{new762}
T. L. Gilbert, A Lagrangian formulation of the gyromagnetic equation of the magnetic field, Phys. Rev. {\bf 100}, 1243 (1955).

\bibitem{new763}
T. L. Gilbert, A phenomenological theory of damping in ferromagnetic materials, IEEE Trans. Magn. {\bf 40}, 3443 (2004).

\bibitem{new77}
J. Xiao, G. E. W. Bauer, K. C. Uchida, E. Saitoh, and S. Maekawa, Theory of magnon-driven spin Seebeck effect, Phys. Rev. B {\bf 81}, 214418 (2010).

\bibitem{new78}
S. K. Kim and Y. Tserkovnyak, Landau-Lifshitz theory of thermomagnonic torque, Phys. Rev. B {\bf 92}, 020410(R) (2015).

\bibitem{new79}
L. Arrachea and F. von Open, Nanomagnet coupled to quantum spin Hall edge: An adiabatic quantum motor, Physica E {\bf 74}, 596-602 (2015).

\bibitem{new83}
J. Fransson, Detection of spin reversal and nutations through current measurements, Nanotechnology {\bf 19}, 285714 (2008).

\bibitem{bodecurrentinduced}
N. Bode, L. Arrachea, G. S. Lozano, T. S. Nunner, and F. von Oppen,  Current-induced switching in transport through anisotropic magnetic molecules, Phys. Rev. B {\bf 85}, 115440 (2012).

\bibitem{new80}
L. Berger, Low-field magnetoresistance and domain drag in ferromagnets, J. Appl. Phys. {\bf 49}, 2156-2161 (1978).

\bibitem{new81}
S. Zhang and Z. Li, Roles of Nonequilibrium Conduction Electrons on the Magnetization Dynamics of Ferromagnets, Phys. Rev. Lett. {\bf 93}, 127204 (2004).

\bibitem{new82}
K. M. D. Hals and A. Brataas, Spin-orbit torques and anisotropic magnetization damping in skyrmion crystals, Phys. Rev. B {\bf 89}, 064426 (2014).

\bibitem{new85}
M. O. A. Ellis, M. Stamenova, and S. Sanvito, Multiscale modeling of current-induced switching in magnetic tunnel junctions using \textit{ab initio}
spin-transfer torques, Phys. Rev. B {\bf 96}, 224410 (2017).

\bibitem{t6}
U. Bajpai and B. Nikoli\'c, Time-retarded damping and magnetic inertia in the Landau-Lifshitz-Gilbert equation self-consistently coupled to electronic time-dependent nonequilibrium Green functions, Phys. Rev. B {\bf 99}, 134409 (2019).

\bibitem{new86}
A. Suresh, U. Bajpai, and B. K. Nikoli\'{c}, Magnon-driven chiral charge and spin pumping and electron-magnon scattering from time-dependent quantum transport combined with classical atomistic spin dynamics, Phys. Rev. B {\bf 101}, 214412 (2020).

\bibitem{new87}
M. Elbracht and M. Potthoff, Accessing long timescales in the relaxation dynamics of spins coupled to a conduction-electron system using absorbing boundary conditions, Phys. Rev. B {\bf 102}, 115434 (2020).

\bibitem{t8}
R. Smorka, M. Thoss, and M. \v{Z}onda, Dynamics of spin relaxation in nonequilibrium magnetic nanojunctions, New. J. Phys. {\bf 26}, 013056 (2024).

\bibitem{4rad}
M. Filipovi\'{c}, Effect of uniaxial magnetic anisotropy on charge transport in a junction with a precessing anisotropic molecular spin, Phys. Rev. B {\bf 111}, 165415 (2025).

\bibitem{Floquet1}
G. Floquet, Sur les \'{e}quations diff\'{e}rentielles lin\'{e}aires \`{a} coefficients p\'{e}riodiques, Ann. Sci. \'{E}cole Normale Sup\'{e}rieure {\bf 12}, 47 (1883).

\bibitem{Floquet2}
J. H. Shirley, Interaction of a Quantum System with a Strong Oscillating Field, PhD Thesis, California Institute of Technology (1963).

\bibitem{Floquet3}
M. Grifoni and P. H\"{a}nggi, Driven quantum tunneling, Phys. Rep. {\bf 304}, 229 (1998).

\bibitem{Floquet4}
B. H. Wu and C. Timm, Noise spectra of ac-driven quantum dots: Floquet master-equation approach, Phys. Rev. B {\bf 81}, 075309 (2010).

\bibitem{Fano2024}
U. Fano, Effects of Configuration Interaction on Intensities and Phase Shifts, Phys. Rev. {\bf 124}, 1866 (1961).

\bibitem{interf2024}
A. E. Miroshnichenko, S. Flach, and Y. S. Kivshar, Fano resonances in nanoscale structures, Rev. Mod. Phys. {\bf 82}, 2257 (2010).

\bibitem{Kittel}
C. Kittel, On the theory of ferromagnetic resonance absorpton, Phys. Rev. {\bf 73}, 155 (1948). 

\bibitem{Godfrin}
C. Godfrin, S. Thiele, A. Ferhat, S. Klyatskaya, M. Ruben, W. Wernsdorfer, and F. Balestro, Electrical Read-Out of a Single Spin Using an Exchange-Coupled Quantum Dot, ASC Nano {\bf 11}, 3984 (2017).

\bibitem{Guo}
B. Wang, J. Wang, and H. Guo, Quantum spin field effect transistor, Phys. Rev. B {\bf 67}, 092408 (2003).

\bibitem{Bruus}
H. Bruus and K. Flensberg, \textit{Many-Body Quantum Theory in Condensed Matter Physics} (Oxford University Press, Oxford, UK, 2004).


\bibitem{Sauret}
O. Sauret and D. Feinberg, Spin-Current Shot Noise as a Probe of Interactions in Mesoscopic Systems, Phys. Rev. Lett. {\bf 92}, 106601 (2004).

\bibitem{Wick2024}
A. Fetter and J. D. Walecka, \textit{Quantum Theory of Many-Particle Systems} (Dover, Mineola, NY, 2003).

\bibitem{Langreth2024}
D. C. Langreth, in \textit{Linear and Nonlinear Electron Transport in Solids}, edited by J. T. Devreese and E. Van Doren (Plenum, New York, 1976).


\end{thebibliography}
\end{document}